\pdfoutput=1
\documentclass[modern]{aastex701}
\makeatletter\protected\expandafter\def\expandafter\@endpbox\expandafter{\@endpbox}\makeatother
\def\today{November 11, 2026}
\usepackage[T1]{fontenc}
\usepackage{newtxtext}
\usepackage{newtxmath}
\usepackage{xcolor}
\newcommand{\okina}{\textquoteleft}
\newcommand{\obj}{A108661}

\newcommand{\refAshort}{30.2158}
\newcommand{\refAsigShort}{0.0004}
\newcommand{\refEshort}{0.0661}
\newcommand{\refIshort}{20.32}

\newcommand{\refChi}{79.42}
\newcommand{\refDof}{102}
\newcommand{\refFrames}{88}

\newcommand{\refNights}{54}

\newcommand{\refArc}{17.22}

\newcommand{\refPsvRows}{88}

\newcommand{\foUsed}{88}

\newcommand{\foDiffMax}{0.37}
\newcommand{\foSigDevPct}{4.6}

\newcommand{\nCandidates}{367}

\newcommand{\refApo}{32.2}
\newcommand{\refPeriShort}{28.2}
\newcommand{\refChiPerDof}{0.78}
\newcommand{\looMean}{1.56}
\newcommand{\looMax}{5.23}
\newcommand{\looOver}{0}
\newcommand{\dynSurvived}{421}
\newcommand{\dynClones}{500}
\newcommand{\dynPct}{84}
\newcommand{\dynPctErr}{2}
\newcommand{\dynOtherLo}{79}
\newcommand{\dynOtherHi}{84}
\newcommand{\dynAmpMyr}{8.7}
\newcommand{\dynAmpKyr}{7.7}
\newcommand{\dynPeriodKyr}{9.4}
\newcommand{\dynAmpRun}{31.5}
\newcommand{\dynPeriodMyr}{9.25}
\newcommand{\dynAmpLast}{25.6}
\newcommand{\dynExitFirst}{1.2}
\newcommand{\dynExitMedian}{3.3}
\newcommand{\dynAmpLastMed}{22.0}
\newcommand{\dynAmpLastLo}{6.9}
\newcommand{\dynAmpLastHi}{31.7}
\newcommand{\dynPerMed}{9.36}
\newcommand{\dynPerLo}{9.23}
\newcommand{\dynPerHi}{9.48}
\newcommand{\dynCenMed}{61.5}
\newcommand{\dynCenLo}{59.7}
\newcommand{\dynCenHi}{63.6}
\newcommand{\dynEmax}{0.092}
\newcommand{\dynEnergy}{1.85}
\newcommand{\dynCtlHalf}{15}
\newcommand{\dynCtlFull}{17}
\newcommand{\dynCtlP}{0.69}
\newcommand{\dynCtlAmpHalf}{29.2}
\newcommand{\dynCtlAmpFull}{29.5}
\newcommand{\knownLfour}{31}
\newcommand{\knownAmpBelow}{5}
\newcommand{\knownAmpMedian}{16}
\newcommand{\knownAmpLo}{3}
\newcommand{\knownAmpHi}{31}
\newcommand{\knownIncAbove}{15}
\newcommand{\knownEAbove}{14}

\shorttitle{A Bright L4 Neptune Trojan in Archival ZTF Images}
\shortauthors{Jongma}

\begin{document}

\title{A Bright L4 Neptune Trojan Found by Stacking Archival ZTF Images}

\author[0009-0005-3744-6238]{Jorrit B. Jongma}
\affiliation{Independent researcher, The Netherlands}
\email{research@jongma.org}

\begin{abstract}
We report a Neptune Trojan found by a shift-and-stack search of archival Zwicky Transient Facility (ZTF) difference images of 2019
October, and recovered in public images of seven other telescopes back to 2008. The orbit uses \refFrames{} positions on \refNights{}
nights from 2008 to 2025, a \refArc-year arc, all measured on our own Gaia DR3 plate solutions. Its heliocentric osculating elements at JD 2460000.5
are $a = \refAshort \pm \refAsigShort$~au,
$e = \refEshort$ and $i = \refIshort$\arcdeg; a fit of the same positions with the independent orbit code Find\_Orb agrees within
0.4$\sigma$ in every element.
The nominal orbit librates about Neptune's leading Lagrange point L4 with a period of \dynPeriodMyr~kyr and an amplitude of
\dynAmpMyr\arcdeg{} over the next Myr, and $\dynPct \pm \dynPctErr$\% of \dynClones{} clones of the orbit remain L4 Trojans for 4.5~Gyr.
Its absolute magnitude is $H_V = 6.96 \pm 0.10$ (linear phase law), or $6.81 \pm 0.06$ on the $H$, $G$ convention of the Minor
Planet Center's list of Neptune Trojans, whose brightest L4 member has $H = 7.26$, so the object is possibly the brightest known L4 Neptune Trojan.
Its color, $g - r = 0.52 \pm 0.05$, lies in the $g - r$ range of the red Neptune Trojans, and for albedos of 0.05--0.15 its diameter is about 140--240~km.
\end{abstract}


\section{Introduction}\label{sec:intro}

Neptune Trojans share Neptune's orbit and librate about its L4 or L5 Lagrange point, 60\arcdeg{} ahead of or behind the
planet. The first was found in 2001 \citep{chiang2003}; the population is mostly dynamically hot (inclinations to tens of degrees)
\citep{sheppard2006,lin2021}, is found at L4 \citep{gerdes2016} and L5 \citep{sheppard2010}, and is thought to have been
captured during Neptune's migration \citep{nesvorny2009,lykawka2009,gomes2016}. It appears several times larger than the Jupiter Trojan population \citep{sheppard2006}.
Its long-term stability has been
mapped with integrations of up to 1~Gyr \citep{dvorak2007} and 34~Myr, checked to 4.5~Gyr along two lines \citep{zhou2009}; unstable members are a possible source of
Centaurs \citep{horner2010}. Their colors resemble those of the Jupiter Trojans \citep{jewitt2018}; a few ultra-red members have since been found
\citep{lin2019,markwardt2023}.
The Minor Planet Center's list of Neptune Trojans \citep{mpcntlist}, last updated 2026 May 30, holds 28 objects at L4 and 7 at L5.
The known Neptune Trojans were found in deep surveys of the outer solar system \citep{chiang2003,sheppard2006,lin2021}, in
wide-field surveys \citep{lin2016} and in the Dark Energy Survey \citep{gerdes2016,bernardinelli2022}; the two brightest listed L4
objects are (530930) 2011~WG$_{157}$, found by Pan-STARRS1 (PS1) in 2011 \citep{lin2016}, and 2015~VU$_{207}$, found in Dark Energy
Survey images of 2015 \citep{mpec2021s63}.

Here we report an L4 Trojan found in archival ZTF data and recovered in the archives of seven other telescopes, with its orbit,
the evidence that all its positions belong to one object, its long-term dynamics and its physical properties. The object has no
Minor Planet Center designation yet; we use our internal identifier \obj.
Our first positions were reported to the Minor Planet Center on 2026 November 9.
We find no earlier report of it in the Minor Planet Center's records, apart from one unlinked tracklet of 2025 November 21 in its
Isolated Tracklet File (Section~\ref{sec:orbit}).
The frames are listed in Appendix~\ref{app:frames}, the measurement details, frame criteria and linkage tests in
Appendices~\mbox{\ref{app:measure}--\ref{app:linkage}}, and the positions are released with the paper (Section~\ref{sec:availability}).

\section{Discovery}\label{sec:discovery}

\obj{} was found in a search of archival ZTF \citep{bellm2019,graham2019,masci2019,dekany2020} $r$-band difference images for
moving objects. The search covered 93 fields on 327 field-nights (255 in 2019, 40 in 2018, 31 in 2020 and 1 in 2024), about
3,640~deg$^2$. The images of each CCD and night were stacked on a GPU along straight tracks at 0--60$''$/h; a tracklet needed a
search S/N (that of the stack along its track) of at least 9 (10 above 5$''$/h, where tracklets of S/N 9--10 only extend
linked objects), and its position and rate came from an
independent PSF fit along the track where that fit was accepted (Appendix~\ref{app:measure}); tracklets whose PSF
fit failed were kept, flagged, at the search's position and rate. Sources planted in the images (150 per CCD) put the 50\% detection limit of the discovery nights at $V \approx 22.2$.
A further $\sim$1,500~deg$^2$ was searched previously with different parameters.
Of the 71,766 tracklets current at 2026 November 1, 11:15 UTC, we inspected image stamps of the 122 that the
automatic match had then not identified with any object of the Minor Planet Center's orbit catalog and observation files (18 had
only possible matches, 16 of them more than 3$'$ away), had not been linked to other nights, moved slower than
10$''$/h and had a search S/N of at least 15 and an accepted PSF fit; these cuts were set before the inspection. A real moving
source is a point source in each third of the night, absent from the static sky, and gone when the exposures are stacked along a
track fixed on the detector. Three real objects remained: \obj, Uranus's moon
Sycorax (the Minor Planet Center files of the match hold no natural satellites) and a fainter object seen on one night of 2024
January; two more were known asteroids whose orbits ($U = 3$) were above the automatic match's limit ($U \leq 2$), and four stayed
unclear.

\begin{figure*}[t!]
\plotone{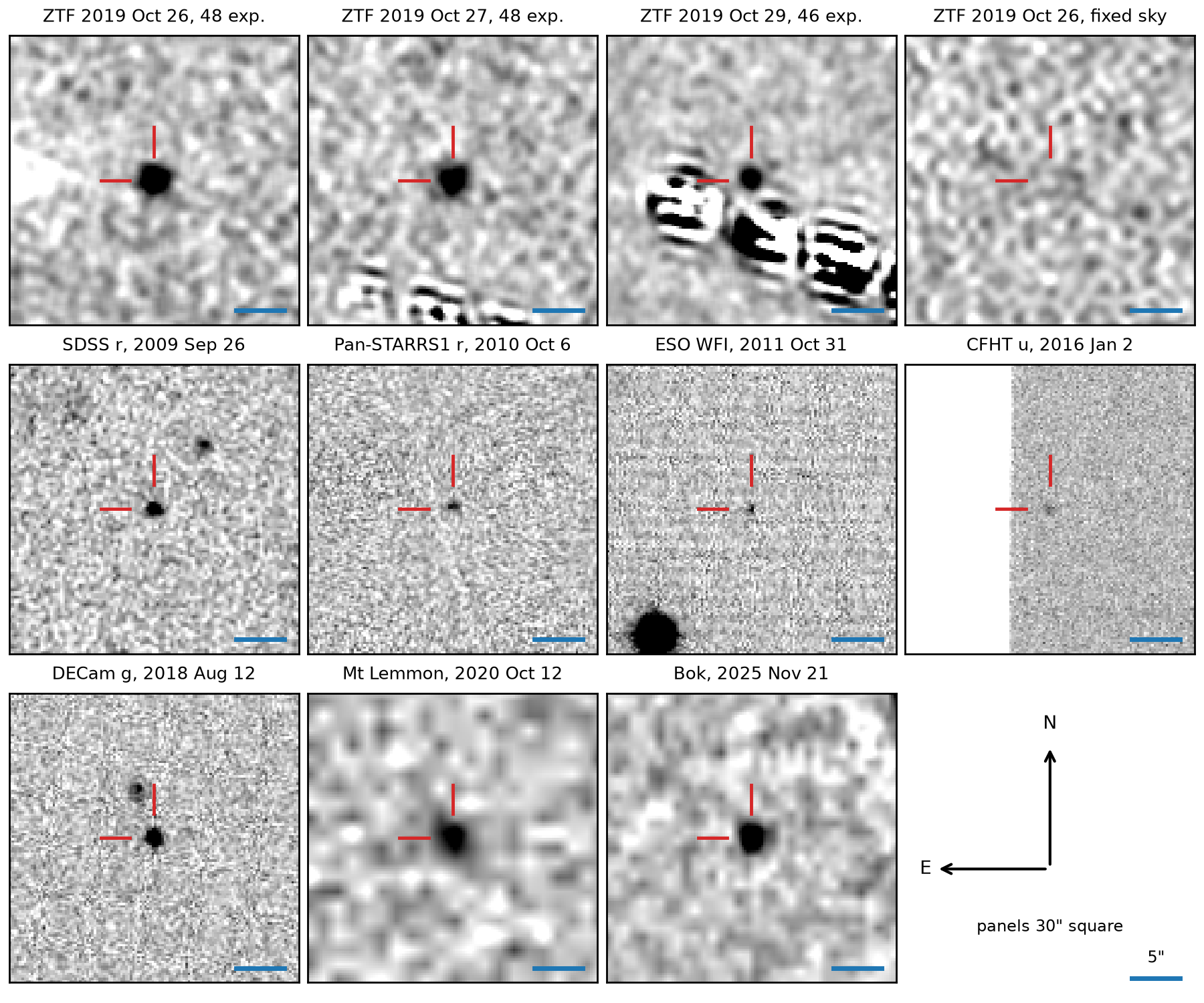}
\caption{\obj{} in ZTF and in the archives. Top row: the ZTF difference images of each discovery night co-added following the
object, and the October 26 images co-added at one fixed sky position, where its light is spread along a 26$''$ track.
Other rows: per telescope the single image (not a difference image) of the frame of the orbit with the highest S/N in
Table~\ref{tab:frames}: SDSS frame-r-008055-2-0059,
PS1 1852.019.wrp.r.55475\_52082, WFI.2011-10-31T02:51:50.142 CHIP4, CFHT 1859823p ccd13, DECam c4d\_180812\_091515\_ooi\_g\_v1,
Mt Lemmon g96\_20201012\_2b\_n27017\_01\_0002, Bok v00\_20251121\_2a\_n26048\_01\_0003. Ticks mark our measured position. Each panel
is 30$''$ square, north up, east left; bar 5$''$. The panels are resampled bilinearly, without smoothing, onto a 0.25$''$ grid with
each archive's header WCS, which differs from our plate solution by up to a few tenths of an arcsecond, so the object can sit
slightly off the ticks; the coarse pixels of Mt Lemmon (1.5$''$) and Bok (0.9$''$) make those panels look smooth. White: outside
the CCD (CFHT).\label{fig:discovery}}
\end{figure*}

It was detected in field 605 (CCD 11, quadrant 4) on 2019 October 26 and 27 (search S/N 17.2 and 15.6) at $r \approx 21.3$; on
October 26 at 08:02~UTC it was at RA $2^{\rm h}32^{\rm m}08\fs59$, Dec $+26\arcdeg36'45\farcs8$, moving at 4.3$''$/h toward
position angle 250\arcdeg, a rate that near opposition places it at about 29~au from the Sun. On October 29 the object lies next to a
star-subtraction residual (Figure~\ref{fig:discovery}); its tracklet failed the automatic PSF fit (and its search S/N, 14.3, was below 15), so it was not among the 122 and
we added it to the inspection by hand; the pipeline's linking later joined it to the other two nights. The three discovery nights have 48, 48 and 46 exposures; the 9 exposures of October 28 are too shallow to
show it.

The three nights alone leave the orbit nearly unconstrained: of 2,000 orbits sampled along the line of variation of a fit to
them, 95\% (with the measured errors) to 99.5\% (with 0.3$''$ errors) are unbound. Table~\ref{tab:chain} shows how the orbit was
narrowed. The searches assumed no orbit class: the first window stacks followed a family of near-circular, prograde orbits
($i < 45\arcdeg$), the season stacks every orbit still allowed by the data. These ZTF stacks of other weeks and seasons found the
object again and narrowed the orbit until single archival frames could be searched: first the DECam frames of 2018, then the
other archives at the places the orbit predicted. A stack finds the object and gives a rough position but is not a measurement:
the orbit of Section~\ref{sec:orbit} rests only on single frames and the ZTF sub-stacks of the discovery nights. Each search was planned from an orbit without the data it searched; six stack searches missed one of
their preset acceptance rules, four of them by a small margin (Appendix~\ref{app:linkage}), and all served only as guides. As a
check of the orbit class, four hypotheses (free orbit, Neptune Trojan, 3:2 and 2:1 resonance) were then searched side by
side with a different code from the first discovery night alone: only the Trojan hypothesis was found in the two 2019 windows
(a marginal free-orbit hit is its own light), and then in every season of 2018--2025 (2022--2025 after re-planning on it;
Appendix~\ref{app:linkage}).

\edef\savedtabcolsep{\the\tabcolsep}\setlength{\tabcolsep}{3pt}
\begin{deluxetable*}{r>{\raggedright\arraybackslash}p{3.5cm}>{\raggedright\arraybackslash}p{1.6cm}>{\raggedright\arraybackslash}p{2.6cm}>{\raggedright\arraybackslash}p{1.5cm}r>{\raggedright\arraybackslash}p{3.0cm}}
\tabletypesize{\scriptsize}
\tablecaption{How the orbit was narrowed. Each step adds new data (a \emph{sub-stack} co-adds one night's consecutive ZTF exposures along
the object's motion; a \emph{window} or \emph{season stack} co-adds the ZTF exposures of weeks or of a whole season along each
trial orbit; a \emph{frame} is one exposure). ``Planned from'' gives the data of the orbit that predicted where to look; that orbit
never contained the data then searched. $\sigma_a$ is the half-width of the central 68\% of the orbit samples (steps 1--5) or the
formal 1$\sigma$ error (steps 7--11); the osculating epochs differ from step to step, so only the widths compare. $\sigma_{\rm pos}$
is, on 2026 December 22, one year after the last measured frame (geocentric), half the central 68\% range of the sampled
positions along their major axis (steps 1--5) or the semi-major axis of the predicted 1$\sigma$ uncertainty ellipse (steps 7--11). Step 1 gives the share of
unbound orbit samples (with each position's measured error, and with 0.3$''$ on each). Step 6 is a check, not a refinement. Step
10 is a fit to frame positions made before the frame criteria and night errors of Section~\ref{sec:data}.
\label{tab:chain}}
\tablehead{\colhead{Step} & \colhead{New data} & \colhead{Kind} & \colhead{Planned from} & \colhead{$\sigma_a$ (au)} &
\colhead{$\sigma_{\rm pos}$ ($''$)} & \colhead{Result}}
\startdata
1 & the three discovery nights, 2019 October & sub-stacks & the blind search (Section~\ref{sec:discovery}) & unbound in 95--99.5\% & $6.5\times10^{4}$ & the orbit is hardly constrained \\
2 & window stacks of the 2019 season before and after the nights & window stacks & the three nights (the second round: also the first round's two windows) & 2.7 & 3100 & found again; bound but wide \\
3 & season stacks of 2018 and 2020; DECam frames of 2018 & season stacks; frames & step 2 (no data of 2018 or 2020) & 0.037 & 32 & narrow enough to search archives; first single-frame detections \\
4 & season stacks of 2021 and 2022; five DECam positions in the fit & season stacks; frames & steps 1--2 and the 2020 season stack (no 2018 data) & 0.0047 & 4.1 & librates about L4 in a 1~Myr integration \\
5 & season stacks of 2023 to 2025 & season stacks & steps 1--4, with 3 times the stack position errors & 0.0044 & 1.2 & predicts the Isolated Tracklet File tracklet of 2025 November 21 (Section~\ref{sec:orbit}) \\
6 & four orbit hypotheses side by side, in the same two 2019 windows & window and season stacks & the first discovery night alone (2022--2025: re-planned on the Trojan solution) & -- & -- & a check: only the Trojan hypothesis is found in the two 2019 windows (a marginal free-orbit hit is its own light); it alone is then searched, and found, in every season of 2018--2025 \\
7 & single-night positions only: ZTF 2019 sub-stacks, DECam 2014 to 2018 & sub-stacks; frames & DECam 2014, 2015: the stack orbit & 0.0037 & 3.1 & an orbit without any stack \\
8 & Pan-STARRS1, 2010 to 2014 & frames & step 7 (no PS1 data) & 0.00085 & 0.33 & the arc reaches back to 2010 \\
9 & SDSS, 2008 to 2009 & frames & step 8 (no SDSS data) & 0.00068 & 0.25 & the arc reaches back to 2008 \\
10 & ESO WFI, CFHT, Mt Lemmon, Bok & frames & steps 7--9 & 0.00035 & 0.072 & every archive on the path \\
11 & every frame re-measured on Gaia DR3; the frame criteria & frames; sub-stacks & no new search & 0.00039 & 0.069 & the orbit of this paper \\
\enddata
\end{deluxetable*}
\setlength{\tabcolsep}{\savedtabcolsep}

\section{Data and measurements}\label{sec:data}

Table~\ref{tab:telescopes} lists the telescopes, their archives and how the object was found in each: SDSS DR17
\citep{york2000,eisenstein2011,gunn1998,gunn2006,abdurrouf2022}, Pan-STARRS1 DR2 \citep{chambers2016,magnier2020c,waters2020,magnier2020b,magnier2020,flewelling2020},
the ESO/MPG 2.2~m Wide Field Imager \citep{baade1999}, DECam \citep{flaugher2015,valdes2014} (all ten frames from the DECam
Legacy Survey), CFHT MegaCam \citep{boulade2003,magnier2004}, ZTF, and the Mt Lemmon 1.5~m, Steward Mt Lemmon 1.0~m and Bok 2.3~m telescopes of the
Catalina Sky Survey \citep{css}, whose frames (PDS Small Bodies Node, \citealt{seaman2022}) were located along the orbit with CATCH \citep{kelley2021}. Of each SDSS drift scan only
the $r$ frame is used, because the bands of a scan are not independent measurements
(Table~\ref{tab:rules}). ZTF contributes seven sub-stacks: the thirds of 2019 October 26 and 27 (16 exposures each) and the whole
night of October 29 (46 exposures), whose thirds do not all give usable fits. All 43 frames of the Steward Mt Lemmon 1.0~m were
taken with non-sidereal tracking. Survey catalog positions are not used.

\textbf{Astrometry.} Every position is our own measurement on our own plate solution of the frame against Gaia DR3
\citep{gaia2016,gaia2023,lindegren2021}, whose stars are moved to the frame's time with their proper motions and parallaxes.
The object's position and flux come from a free fit with the frame's own PSF; forced fits at blank spots give the noise
scale below.
ZTF and DECam positions are measured on difference images, the others on the single image; the
exposures of a ZTF sub-stack are fitted together along a track that follows the object's motion. Appendix~\ref{app:measure} gives
the plate solutions, PSFs and settings per telescope.

\textbf{Noise-calibrated S/N.} A fit's formal error need not describe the real noise of an image (resampled and difference images
have correlated noise), so the S/N is calibrated on the image itself. The noise-calibrated S/N of a frame is its free-fit S/N divided
by the spread (standard deviation) of the S/N of forced fits at 300 blank-sky spots in the same image (200 for DECam; for a ZTF
sub-stack the 250--294 source-free ones of 300 blank tracks, Appendix~\ref{app:measure}); for PS1 and ZTF the spread counts as at least 1, and CFHT and the
Mt Lemmon telescopes must pass with and without that floor. Its false-alarm fraction is the share of those spots at which a free
fit, made and accepted the same way, reaches the frame's own free-fit S/N.

\textbf{Errors.} A frame's error is its fit error scaled for the noise excess of blank sky and the measurement's own error scale,
plus per-frame terms (Appendix~\ref{app:measure}). The frames of a night give one night position, whose error adds the telescope's
systematic floor and any offset uncertainty once (Table~\ref{tab:telescopes}), measured on known asteroids of the same telescope,
mostly against JPL Horizons \citep{horizons} (for DECam partly on frames of other pipeline versions; Appendix~\ref{app:measure}),
for ZTF on 5,996 known-asteroid sub-stacks.
Differential chromatic
refraction is not corrected; its effect on the known asteroids is part of what the floors measure.

\textbf{Timing.} Times are mid-exposure UTC from each frame's own header, never from an archive listing. For an SDSS drift scan the time is that of the
object's predicted pixel row.

\textbf{Frame selection.} A frame enters the orbit only if its noise-calibrated S/N is at least 5, its false-alarm fraction at
most 1\% and no cataloged neighbor disturbs its fit, and, where tested (Table~\ref{tab:rules}), if no fixed source seen on other
nights lies under it, its pixels at the object are clean, its free fit does not stop at its position bound and lies within 1.5$''$ of the predicted place, no Gaia star and
no blend (a fit much brighter than the object) disturbs it, and the camera tracked at the sidereal rate. The seven ZTF sub-stacks
are a fixed list, not tested by the S/N, false-alarm and fit criteria; their noise-calibrated S/N is 8.95--14.33, and no free fit
on any of their 300 blank tracks reaches it. A night whose frames disagree ($p < 0.001$) would be removed; none is. No frame is
dropped for its residual alone; outliers are checked in the images (Appendix~\ref{app:rules}). Table~\ref{tab:rules} gives every criterion with its threshold and the frames it removes, and
Table~\ref{tab:frames} the verdict on each of the \nCandidates{} candidate frames.

\edef\savedtabcolsep{\the\tabcolsep}\setlength{\tabcolsep}{4pt}
\begin{deluxetable*}{clllcrrrll}
\tabletypesize{\scriptsize}
\tablecaption{The telescopes, in the order their archives were searched (Step: the step of Table~\ref{tab:chain}; the archives of step 10 were searched together). Years: of the frames used (Steward 1.0~m: of its candidates). Candidates are the frames of
Table~\ref{tab:frames} (the SDSS frames removed by ``one per exposure'' are not counted); ``used'' are those in the orbit. The floor is the systematic
error (RA$^*$/Dec, where RA$^*$ is the right-ascension offset times $\cos\delta$) added once per night; it is measured on known
asteroids in data of the same telescope against JPL Horizons, as the quadrature term that brings their $\chi^2$/dof to 1 or, where
that is poorly constrained, its 84\% bootstrap value (exceptions to the samples and to the rule in Appendix~\ref{app:measure}), so a floor of 0.000 means that the asteroids' scatter on that axis is already explained by their measured errors. DECam's rows are split by search step; it has one floor per year.
Found: \emph{blind}, found by the search of Section~\ref{sec:discovery}, with no orbit; \emph{searched}, found by a source search
along the paths of the sampled orbits of Table~\ref{tab:chain} step 2 (Appendix~\ref{app:linkage}), not measured at one predicted
place; \emph{predicted}, measured at the place predicted by an orbit fitted without those data. Archives: the SDSS Science Archive Server, MAST (Pan-STARRS1 DR2), the ESO archive, the NOIRLab
Astro Data Archive, the Canadian Astronomy Data Centre, IRSA, and the PDS Small Bodies Node (found with CATCH).\label{tab:telescopes}}
\tablehead{\colhead{Step} & \colhead{Telescope} & \colhead{Code} & \colhead{Archive} & \colhead{Years} & \colhead{Candidates} & \colhead{Used} &
\colhead{Nights} & \colhead{Floor ($''$)} & \colhead{Found}}
\startdata
1 & Palomar 48-inch ZTF & I41 & IRSA & 2019 & 7 & 7 & 3 & 0.075/0.035 & blind \\
3--4 & Blanco 4 m DECam & W84 & NOIRLab & 2018 & 7 & 7 & 5 & 0.018/0.018 & 5 searched \\
 & & & & & & & &  & 2 predicted \\
7 & Blanco 4 m DECam & W84 & NOIRLab & 2014--2015 & 3 & 3 & 2 & 2014: 0.046/0.048 & predicted \\
 & & & & & & & & 2015: 0.020/0.028 &  \\
8 & Pan-STARRS1 1.8 m & F51 & MAST (DR2) & 2010--2014 & 32 & 12 & 8 & 0.077/0.000 & predicted \\
9 & SDSS 2.5 m & 645 & SDSS DR17 & 2008--2009 & 3 & 3 & 3 & 0.039/0.020 & predicted \\
10 & ESO/MPG 2.2 m WFI & 809 & ESO & 2011 & 2 & 2 & 1 & 0.038/0.006 & predicted \\
10 & CFHT 3.6 m MegaCam & 568 & CADC & 2016 & 2 & 1 & 1 & 0.030/0.000 & predicted \\
10 & Mt Lemmon 1.5 m & G96 & PDS SBN & 2019--2025 & 264 & 49 & 30 & 0.032/0.039 & predicted \\
10 & Bok 2.3 m & V00 & PDS SBN & 2025 & 4 & 4 & 1 & 0.036/0.013 & predicted \\
10 & Steward Mt Lemmon 1.0 m & I52 & PDS SBN & 2021--2023 & 43 & 0 & 0 & -- & -- \\
\enddata
\end{deluxetable*}
\setlength{\tabcolsep}{\savedtabcolsep}

\section{Orbit}\label{sec:orbit}

\begin{figure*}
\plotone{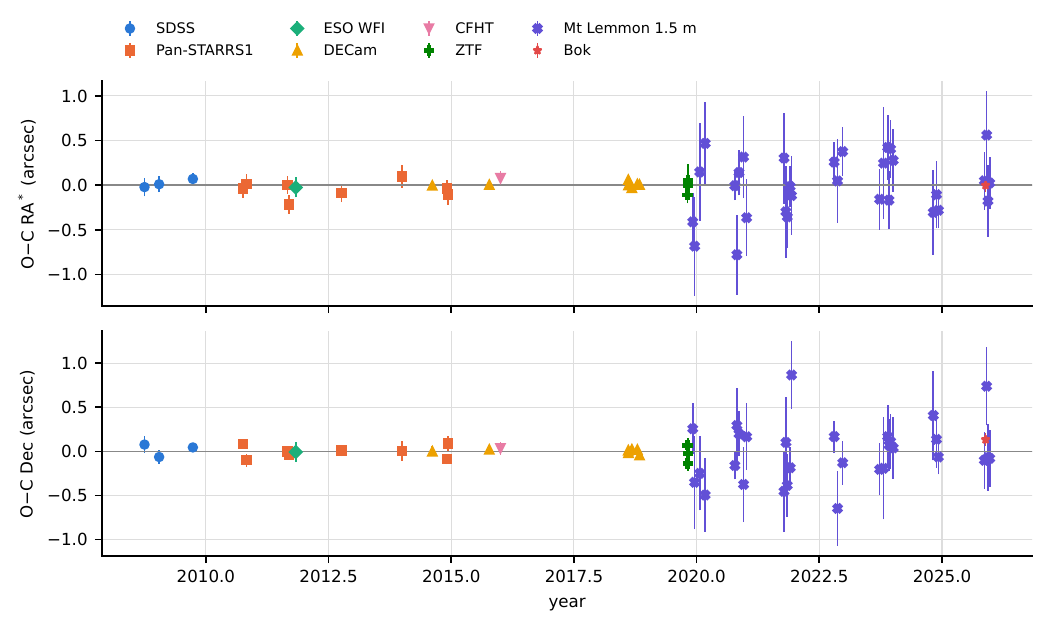}
\caption{Night residuals (observed minus computed) of the orbit in RA$^*$ (defined in Table~\ref{tab:telescopes}) and Dec, per
telescope, with the night errors.\label{fig:residuals}}
\end{figure*}

The orbit is a weighted least-squares fit with an $n$-body model: the Sun, the planets, the Moon and Pluto from JPL DE440
\citep{park2021}, 16 massive asteroids and the Sun's first post-Newtonian term. It fits one position per night and telescope, the
weighted mean of the night's frame residuals with the errors of Section~\ref{sec:data}; the covariance is the formal one, not
scaled. The computed positions are astrometric (ICRS, with the light time iterated) as seen from each telescope's site
(Appendix~\ref{app:measure}), as are the measured ones through their Gaia DR3 plates.
Table~\ref{tab:elements} gives the elements and Figure~\ref{fig:residuals} the residuals: the osculating perihelion and aphelion
distances are \refPeriShort{} and \refApo~au.

The fit has $\chi^2 = \refChi$ for \refDof{} degrees of freedom (\refNights{} night positions $\times$ 2 coordinates $-$ 6
parameters; \refChiPerDof{} per degree of freedom). Predicted by fits without them, the nights lie at a mean squared Mahalanobis
distance (night error plus prediction covariance) of $d^2 = \looMean$ (2 is expected; the largest is \looMax), and \looOver{} of the
\refNights{} nights lie beyond 5.99, the 95\% point of $\chi^2$ for 2 degrees of freedom (about 2.7 expected). Both numbers suggest
that the errors are, if anything, slightly too large (by about 13\% in $\sigma$), so the formal covariance is conservative. The stock
Find\_Orb \citep{findorb}, fitted to the same \foUsed{} frames (DE440; Mercury to Pluto and the Moon as perturbers, no
asteroids; each frame weighted by its error, floor and any offset uncertainty, which count once per frame, not once per night),
rejects no frame and agrees with every element within \foDiffMax$\sigma$ and with every $\sigma$ within
\foSigDevPct\% (Table~\ref{tab:elements}). The predicted position is good to 0.07--0.10$''$ $\times$ 0.03$''$ (1$\sigma$) from 2026 November through 2027.

\textbf{Data not in the fit.} The orbit predicts the positions of the 14 ZTF half-season stacks of 2018 and 2020--2025 and of the
4 halves of the two 2019 window stacks, built along the trial orbits of the Trojan hypothesis (Table~\ref{tab:chain}, step 6),
not along this orbit: their rms offset is 0.39$''$, within 0.77$\sigma$ of their 1$''$ errors each
($\chi^2 = 2.7$ for 36 degrees of freedom; Table~\ref{tab:other}); the 1$''$ errors are generous, so this is a loose check. The four
positions of the Isolated Tracklet File tracklet C1M7NRP, measured by another observer on the Bok images of 2025 November 21 that
also give our Bok night, lie 0.05--0.12$''$ from the prediction along the motion and 0.06--0.49$''$ across it, the across-track offset
growing through the 24-minute tracklet; since they come from the same images, they test the measurement, not the orbit. Before we
measured those images, the orbit of the peak of the 2025 ZTF season stack had predicted the tracklet to within 0.70--0.79$''$; we knew
of the tracklet when that search was run, but its trial orbits had been fixed before we found it. Two single ZTF nights, 2018 November 15 and 2020 November 16, are not in the fit: they were found after the
orbit was fixed and reuse images already in the season stacks; both lie within 1.3$\sigma$ of the prediction on each axis.

\textbf{Sensitivity.} Table~\ref{tab:sensitivity} refits the orbit with one analysis choice changed at a time. No choice moves an
element by more than 0.88$\sigma$ (the node; $a$, $e$, $i$ and $q$ by at most 0.77$\sigma$; both without the Bok night) or
the predicted position of 2026 December 22 by more than 0.045$''$ (a Mahalanobis distance of 1.0$\sigma$ in its 0.069$''$ $\times$
0.031$''$ ellipse). The Bok night, whose weight is one of the choices, has the most leverage on $i$. Re-run from the orbit instead of the earlier
predictions where the fits started, the 1.5$''$ and fit-area tests change no verdict (Appendix~\ref{app:measure}).

\begin{deluxetable}{lllrr}
\tabletypesize{\footnotesize}
\tablecaption{The orbit: osculating heliocentric elements (ecliptic and equinox J2000) at JD 2460000.5 TT (2023 February 25.0),
with formal 1$\sigma$ errors, fitted to 88 frames on 54 nights ($\chi^2 = 79.4$ for 102 degrees of freedom).
The Find\_Orb column is the stock Find\_Orb (version 2026-09-21) fitted to the same 88 frames
(Section~\ref{sec:orbit}); with its 3$\sigma$ outlier rejection it gives the same elements and rejects no frame. $\Delta$ is
Find\_Orb minus this paper in units of this paper's $\sigma$; ``ratio'' is Find\_Orb's $\sigma$ over ours. The barycentric state and
its covariance are in the released orbit file.\label{tab:elements}}
\tablehead{\colhead{Element} & \colhead{This paper} & \colhead{Find\_Orb} & \colhead{$\Delta/\sigma$} & \colhead{Ratio}}
\startdata
$a$ (au) & 30.21584 $\pm$ 0.00039 & 30.21585 $\pm$ 0.00039 & +0.03 & 0.998 \\
$e$ & 0.0661399 $\pm$ 0.0000035 & 0.0661402 $\pm$ 0.0000035 & +0.08 & 1.015 \\
$i$ (deg) & 20.324656 $\pm$ 0.000015 & 20.324661 $\pm$ 0.000016 & +0.36 & 1.034 \\
$q$ (au) & 28.21737 $\pm$ 0.00028 & 28.21737 $\pm$ 0.00028 & +0.01 & 1.001 \\
$\Omega$ (deg) & 254.387887 $\pm$ 0.000036 & 254.387874 $\pm$ 0.000038 & -0.36 & 1.046 \\
$\omega$ (deg) & 214.9158 $\pm$ 0.0079 & 214.9156 $\pm$ 0.0079 & -0.03 & 0.997 \\
$M$ (deg) & 306.8660 $\pm$ 0.0077 & 306.8662 $\pm$ 0.0076 & +0.03 & 0.997 \\
\enddata
\end{deluxetable}

\begin{deluxetable}{p{9.0cm}cc}
\tabletypesize{\footnotesize}
\tablecaption{Sensitivity of the orbit to the analysis choices. Each row is a refit with one choice changed; where a row covers
two or more settings, the largest shift is given. The shift of an element ($a$, $e$, $i$ or $q$) is in units of its $\sigma$;
the last column is the shift of the predicted position on 2026 December 22 (Palomar), whose 1$\sigma$ ellipse is
0.069$''$ by 0.031$''$ (the geocentric ellipse of
Table~\ref{tab:chain}, step 11, is the same to within 0.001$''$). The neighbor row relaxes the catalog part of that criterion; the
four Mt Lemmon frames of 2022 December 24 that fail only the neighbor criterion stay out, because they also fail its static-sky part
(a fixed source at their position in the frames of other nights).
\label{tab:sensitivity}}
\tablehead{\colhead{Changed} & \colhead{Largest shift ($\sigma$)} & \colhead{2026 position ($''$)}}
\startdata
PS1, DECam and SDSS floors (upper bounds, from per-frame scatter) set to zero & 0.30 ($i$) & 0.020 \\
Bok 2025 frame 4 left out (the lowest-S/N Bok frame) & 0.07 ($i$) & 0.006 \\
Neighbor criterion, catalog part, relaxed: its four frames restored (Mt Lemmon 2023 Oct 10, 2025 Nov 12) & 0.12 ($i$) & 0.020 \\
DECam 2014/2015 corrected for their known-asteroid offsets & 0.26 ($e$) & 0.013 \\
DECam 2014/2015 frame times shifted by $\pm 7$~s & 0.03 ($e$) & 0.002 \\
PS1 floor 0.016$''$ or 0 on both axes & 0.18 ($a$, $q$) & 0.011 \\
Telescopes' $\chi^2$ balanced (ZTF errors $\times 1.33$, alone or with PS1 $\times 0.78$) & 0.11 ($a$) & 0.008 \\
PS1 frames selected by the forced-fit instead of the free-fit S/N & 0.22 ($i$) & 0.004 \\
Bok night: common term of 0.072$''$ on both axes & 0.35 ($i$) & 0.015 \\
Bok night: common term of 0.151$''$ on both axes & 0.60 ($i$) & 0.030 \\
All 49 Mt Lemmon frames left out & 0.09 ($e$) & 0.007 \\
The Bok night left out & 0.76 ($i$) & 0.044 \\
\enddata
\end{deluxetable}

\section{Linkage and identity}\label{sec:identity}

Five results tie the \refNights{} nights to one object.
\begin{enumerate}\setlength{\itemsep}{0pt}
\item Apart from the three discovery nights, every archival detection was found along the paths of orbits fitted without its own pixels
(Tables~\ref{tab:chain} and~\ref{tab:telescopes}), and one orbit fits all nights (Section~\ref{sec:orbit}).
\item The DECam 2018 group passed a false-link Monte Carlo on the real search: in 0 of 400 trials did random candidates link as
well as the real group (95\% upper limit 0.0075), while with the real candidates added 392 of 400 did.
\item The sky is empty where the object was: images of other dates show nothing at the 23 PS1 detection spots (forced S/N $-2.2$
to $+1.3$), 17 of which have no catalog detection on any other date, and the SDSS spots are empty in the deeper Legacy Surveys Data
Release 10 catalog (DR10; no source within 6$''$).
\item No known object matches: along the orbit on every day from 2004 to 2026, local copies of the Minor Planet Center's
observation files hold only two sets of positions within 10$''$ of the prediction, the Isolated Tracklet File tracklet of
Section~\ref{sec:orbit} and three Mt Lemmon positions of the main-belt asteroid (220450) 2003~YW$_{111}$, which passed 7--9$''$ away
on 2019 December 17 and shifts our position in the one fitted frame of that night by 0.013$''$ (0.025$\sigma$).
\item PS1's own pipeline detected the object on 11 nights of 2010--2014 (23 detections, 0.06--0.82$''$ from the place predicted by an
orbit without PS1 data), but no observation in our copies of the Minor Planet Center's observation files (Appendix~\ref{app:linkage})
lies within 300$''$ and 1~day of them: these detections were never linked. Our own measurements pass the frame criteria on 7 of those
nights and one further night without a pipeline detection, which gives the 8 PS1 nights of the orbit.
\end{enumerate}

\section{Dynamics}\label{sec:dynamics}

We integrated \dynClones{} clones of the orbit (the nominal state and 499 drawn from the fit's covariance; the starting states
are released, Section~\ref{sec:availability}) with the four giant planets for 4.5~Gyr, with REBOUND 5.2.2 \citep{rein2012} and the
WHFast integrator \citep{wisdom1991,rein2015} with symplectic corrector of order 11 \citep{wisdom1996}, at a 1~yr step, the inner
planets' mass added to the Sun (Appendix~\ref{app:measure}). A clone is counted as lost when $\phi = \lambda - \lambda_{\rm N}$,
sampled every 600~yr, leaves (0\arcdeg, 180\arcdeg). The step samples Jupiter's 12~yr orbit about 12 times; the planets' relative
energy drift between the stored records of 1~Myr and 4.5~Gyr is at most $\dynEnergy\times10^{-7}$, and 20 clones rerun at a 0.5~yr
step keep \dynCtlHalf{} of 20, against \dynCtlFull{} of the same 20 at 1~yr (Fisher's exact test, $p = \dynCtlP$).

\begin{figure*}
\plotone{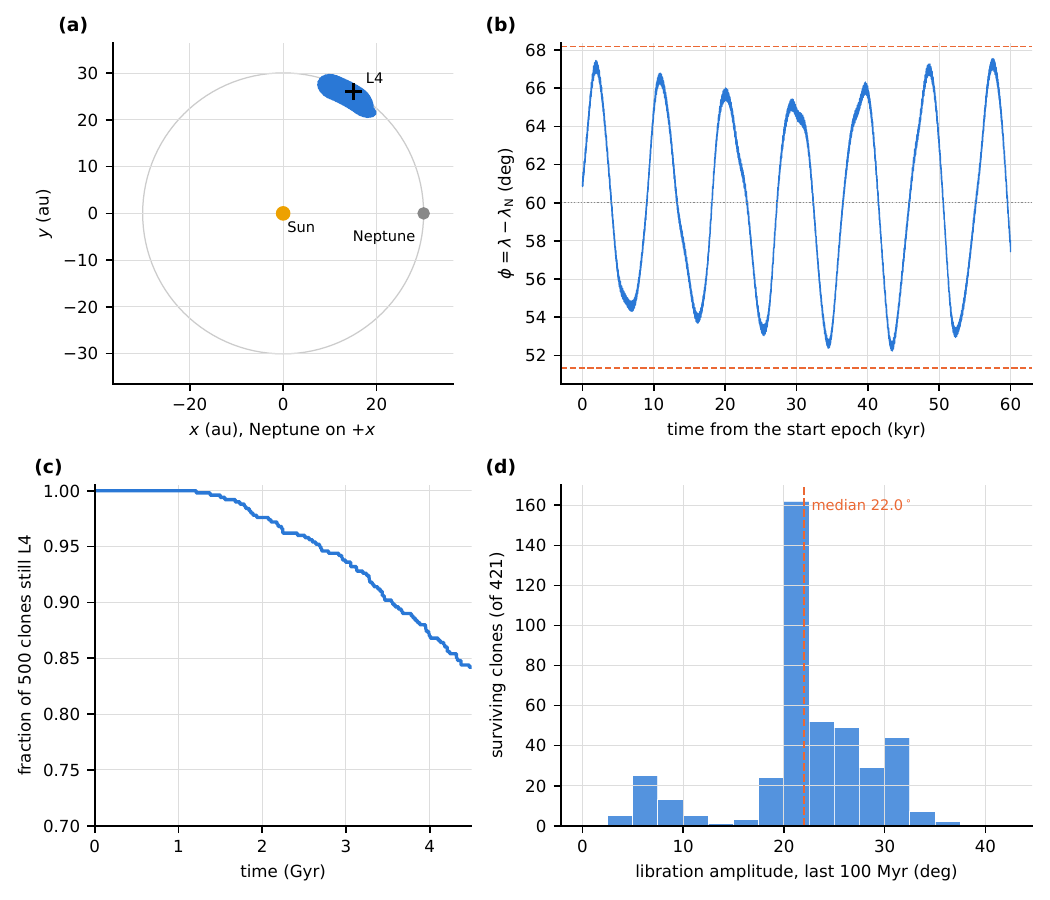}
\caption{Dynamics. (a) The nominal orbit over its first 60~kyr, in heliocentric ecliptic coordinates rotated so that Neptune stays
on the $+x$ axis; the path librates about L4 (cross, 60\arcdeg{} ahead of Neptune; the gray circle has Neptune's mean distance), and
the width of the band comes from the eccentricity (distance from the Sun 28.1--32.1~au over these 60~kyr) and the inclination (the positions are
projected onto the ecliptic). (b) $\phi = \lambda - \lambda_{\rm N}$ of the same orbit every 10~yr; the dashed lines are the minimum
and maximum of $\phi$ over the first Myr in samples every 600~yr (half range 8.4\arcdeg; 8.7\arcdeg{} in the samples every 10~yr),
and the dotted line is 60\arcdeg. (c) Fraction of the
\dynClones{} clones still L4 Trojans versus time. (d) Libration amplitude of the \dynSurvived{} surviving clones in their last
100~Myr (half the full range of $\phi$), with the median.\label{fig:dynamics}}
\end{figure*}

Of the \dynClones{} clones, \dynSurvived{} ($\dynPct \pm \dynPctErr$\%) stay L4 Trojans for 4.5~Gyr (Figure~\ref{fig:dynamics}c).
None leaves in the first Gyr; the others leave between \dynExitFirst{} and 4.5~Gyr (median \dynExitMedian~Gyr), after their amplitude
has slowly grown. The nominal orbit stays for the whole run at the 1~yr step (at 0.5~yr it leaves after 2.7~Gyr). It librates about L4
with a period of \dynPeriodMyr~kyr and an amplitude (half the range of $\phi$) of \dynAmpMyr\arcdeg{} over its first Myr
(Figure~\ref{fig:dynamics}a, b); over 4.5~Gyr the mean period is \dynPeriodKyr~kyr and the half range \dynAmpRun\arcdeg{}
(\dynAmpLast\arcdeg{} in the last 100~Myr); the run's amplitudes, here and below, come from
the samples every 600~yr and are slightly low (by 3\% over the first Myr). The surviving clones have a median amplitude of \dynAmpLastMed\arcdeg{} in their last 100~Myr
(5--95\%: \dynAmpLastLo--\dynAmpLastHi\arcdeg; Figure~\ref{fig:dynamics}d); they librate with periods of
\dynPerLo--\dynPerHi~kyr (5--95\%; median \dynPerMed~kyr) about centers of \dynCenLo--\dynCenHi\arcdeg{} (median
\dynCenMed\arcdeg), and the largest 600-yr sample of their eccentricity is \dynEmax{} (Appendix~\ref{app:measure}). Single clones can change fate with tiny changes of the orbit or the step (8 of the 20 control
clones did), so only the ensemble is meaningful; together with three runs from earlier fits with fewer frames (38 to 81), the four runs gave
\dynOtherLo--\dynOtherHi\%, a spread their counting errors explain ($p = 0.09$).

Over its first 100~kyr the nominal orbit's amplitude is \dynAmpKyr\arcdeg. Our 100~kyr integrations of the Minor Planet Center's
catalog orbits classify \knownLfour{} known objects as L4 Trojans in all seven clones (Figure~\ref{fig:known}): the 28 of the
list and 2012~UW$_{177}$, 2020~VK$_{45}$ and 2020~VN$_{45}$, which the list does not include. Their amplitudes range from \knownAmpLo\arcdeg{} to \knownAmpHi\arcdeg{}
(median \knownAmpMedian\arcdeg) and only \knownAmpBelow{} are smaller; its inclination and eccentricity lie near the middle of
theirs (\knownIncAbove{} of \knownLfour{} have a larger $i$, \knownEAbove{} a larger $e$).

\section{Physical properties}\label{sec:physical}

\textbf{Photometry.} We measured the brightness on the \refFrames{} frames of the orbit with the same fits as the positions, each
telescope calibrated to the SDSS $g$, $r$, $i$ system on its own stars or catalog (Appendix~\ref{app:measure}); 84 frames carry a
calibrated magnitude. Near the S/N threshold of 5 the selection keeps the frames in
which noise made the object brighter: forced photometry of the 49 Mt Lemmon frames of the orbit, most of them near S/N 5, gives $H$
0.21~mag brighter than that of all 122 usable Mt Lemmon frames of the same 30 nights. We therefore take $H$ only from data sets whose frames
are all far above the threshold (noise-calibrated S/N $\geq 8.6$): ZTF, SDSS, DECam and the one PS1 $r$ frame.

\textbf{Absolute magnitude.} Table~\ref{tab:photsize} gives $H$ per band from the night means of the released photometry
(Section~\ref{sec:availability}), fitted with a linear phase slope $\beta$ and a night-to-night scatter $s_{\rm int}$. The 11 $r$
nights cover phase angles 0.40--1.84\arcdeg{} and show no significant phase slope:
$\beta = +0.025 \pm 0.074$~mag/deg, and 0.15~mag/deg, the mean $V$ slope of the trans-Neptunian objects of \citet{rabinowitz2007}, is 1.7$\sigma$ ($r$) to
2.6$\sigma$ ($r$ and $g$ jointly) away; phase slopes of distant objects vary widely, from 0.0 to 0.4~mag/deg below 2\arcdeg{}
in that study, with a median of
0.10~mag/deg for 110 trans-Neptunian objects and Centaurs \citep{alvarezcandal2016}. At the fitted slope the data sets give ZTF
$6.697 \pm 0.035$ (3 nights), DECam $r$ $6.765 \pm 0.043$ (4), SDSS $6.789 \pm 0.118$ (3) and PS1 $r$ $6.686 \pm 0.125$ (1), within
1.2$\sigma$ of each other. We adopt $H_r = 6.75 \pm 0.10$: the fit error of 0.07~mag, which includes the uncertainty of the fitted
slope, widened for the choice of phase law ($\beta = 0$ gives 6.77, the $H$, $G$ function with $G = 0.15$ gives 6.67) and for the
calibrations. DECam $g$ alone gives $H_g = 7.31 \pm 0.04$ ($\beta = 0$). With $V - r = 0.21$ for $g - r = 0.52$ \citep[Table
6]{tonry2012}, $H_V = 6.96 \pm 0.10$. The linear law is adopted because its slope is measured; the $H$, $G$ function with $G = 0.15$
imposes a steeper effective slope of about 0.11~mag/deg over these phase angles.

\textbf{Color.} The two same-night DECam pairs, 2.5 and 2.2~min apart, give $g - r = 0.523 \pm 0.070$ (2015 October 11) and
$0.512 \pm 0.073$ (2018 November 2), mean $0.518 \pm 0.051$; the $g$ and $r$ frames of the same SDSS drift scans give $0.551 \pm 0.080$,
and their $i$ frames $r - i = 0.197 \pm 0.077$. ZTF colors are not used: the object is at or below the detection limit of single ZTF exposures, and
season means combine different nights, so night-to-night changes of the object may enter the color. With
$g - r = 0.52$ the object lies in the $g - r$ range of the red Neptune Trojans of \citet{markwardt2023} (0.31--0.71; their classes are
defined on $g - r$ and $r - i$ together) and is bluer than all their ultra-red members.

\begin{table*}
\centering
\caption{Absolute magnitude and size. Top: $H$ per band from the night means, at the fitted linear phase slope $\beta$ and at two
fixed slopes (mag/deg), with the fitted night-to-night scatter $s_{\rm int}$ (mag; of the fitted-slope fit, for the $g$ row of the
$\beta = 0$ fit). Only the first row is adopted for $H_r$; the
Mt Lemmon row shows the S/N selection bias. Bottom: $H_V$ and the diameter (km; range for $H_V \pm 1\sigma$) for four assumed
geometric albedos $p$.\label{tab:photsize}}
\scriptsize
\edef\savedtabcolsep{\the\tabcolsep}\setlength{\tabcolsep}{4pt}
\begin{tabular}{>{\raggedright\arraybackslash}p{3.6cm}cccccc}
\hline\hline
Set & Nights & $\beta$ (fitted) & $H$ ($\beta$ fitted) & $H$ ($\beta=0$) & $H$ ($\beta=0.15$) & $s_{\rm int}$ \\
\hline
$r$: ZTF + SDSS + DECam $r$ + PS1 $r$ (adopted) & 11 & $+0.025 \pm 0.074$ & $6.746 \pm 0.069$ & $6.767 \pm 0.034$ & $6.647 \pm 0.039$ & 0.088 \\
$r$ + DECam $g$ (one slope, $g$ offset free) & 14 & $+0.012 \pm 0.053$ & $6.756 \pm 0.053$ & -- & -- & 0.081 \\
$g$: DECam $g$ only & 3 & -- & -- & $7.310 \pm 0.043$ & $7.153 \pm 0.077$ & 0.057 \\
$r$: Mt Lemmon (unfiltered; biased, see text) & 30 & $+0.027 \pm 0.088$ & $6.439 \pm 0.086$ & $6.462 \pm 0.039$ & $6.329 \pm 0.040$ & 0.134 \\
\hline
\multicolumn{2}{l}{$H_V$ convention} & $H_V$ & $p = 0.05$ & $p = 0.08$ & $p = 0.10$ & $p = 0.15$ \\
\hline
\multicolumn{2}{l}{ours: linear phase law (adopted)} & $6.96 \pm 0.10$ & 241 (231--253) & 191 (182--200) & 171 (163--179) & 139 (133--146) \\
\multicolumn{2}{l}{$H$, $G$ with $G = 0.15$, list convention} & $6.81 \pm 0.06$ & 259 (251--266) & 204 (199--210) & 183 (178--188) & 149 (145--153) \\
\hline
\end{tabular}
\setlength{\tabcolsep}{\savedtabcolsep}
\end{table*}

\begin{figure*}
\plotone{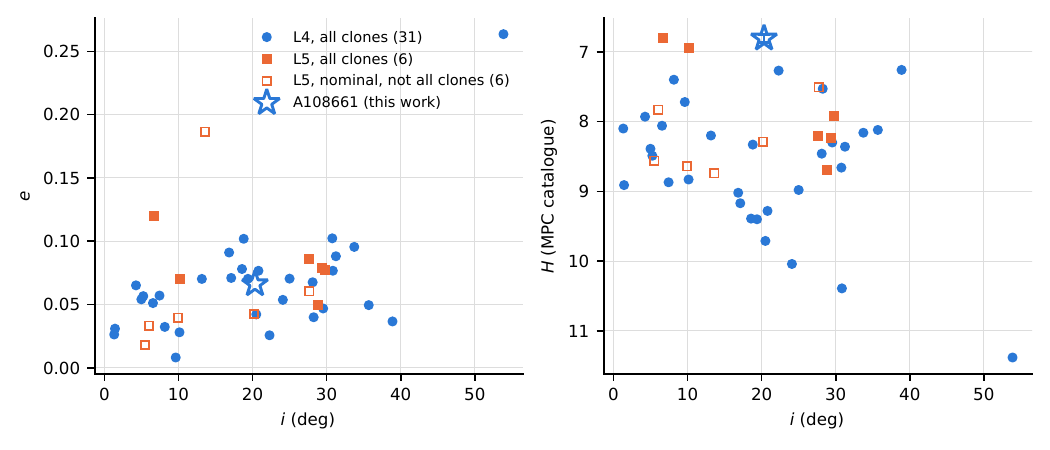}
\caption{Eccentricity (left) and absolute magnitude (right) against inclination for the known Neptune co-orbitals that our 100~kyr
integrations of the Minor Planet Center's orbit catalog (file of 2026 November 4) classify as L4 Trojans in all seven clones
(\knownLfour{} objects; no object is L4 on its catalog orbit without being so in all seven) or as L5 Trojans (filled: all seven
clones; open: the catalog orbit but not every clone). The clones are the catalog orbit and six with $a$ shifted by $\pm$0.5,
$\pm$1 and $\pm$2$\sigma_a$, $\sigma_a$ estimated from the catalog's uncertainty parameter $U$ (at most 0.5~au). Their elements are the
catalog's osculating elements at JD 2461200.5 and their $H$ the catalog's; \obj{} (star) is shown with the elements of the orbit
at JD 2460000.5 and, in the right panel, with $H_V = 6.81 \pm 0.06$ on the catalog's $H$, $G$ convention (our own value is $6.96 \pm 0.10$, Section~\ref{sec:physical}).\label{fig:known}}
\end{figure*}

\textbf{Size and rank.} The diameter $D = 1329~{\rm km}\, p^{-1/2}\, 10^{-H_V/5}$ \citep{cneossize} is 139--241~km for albedos
$p = 0.05$--0.15 (149--259~km on the list's $H$; Table~\ref{tab:photsize}). No Neptune Trojan albedo has been measured
\citep{markwardt2025}. The spectral slope of the DECam color, $4 \pm 3$\% per 100~nm, is closer to the dark neutral trans-Neptunian objects
of \citet{lacerda2014} (neutral on that scale; red among the Neptune Trojans of \citealt{markwardt2023}), whose albedos cluster near
0.05, than to their bright red ones, which favors the larger diameters; Jupiter Trojans have a mean albedo of $0.07 \pm 0.03$ \citep{grav2011}. The Minor Planet
Center's list gives $H$ as $V$ magnitudes at zero phase in the $H$, $G$ system \citep{bowell1989} with $G = 0.15$, from magnitudes
converted to $V$ with fixed offsets ($r \rightarrow V$: $+0.14$~mag; \citealt{mpcbandconv}) and without uncertainties. On that convention our $r$ nights give
$H_V = 6.81 \pm 0.06$.
The brightest listed L4 objects, 2015~VU$_{207}$ ($H = 7.26$) and (530930) 2011~WG$_{157}$ (7.27), are
0.45 and 0.46~mag fainter, so no listed L4 Trojan is brighter; of all 35 listed Neptune Trojans only 2013~KY$_{18}$, at L5 ($H = 6.80$), is as bright. Against our own
$H_V = 6.96$ instead, the two L4 objects are 0.30 and 0.31~mag fainter, while 2013~KY$_{18}$ is 0.16~mag brighter, as is, by 0.02~mag,
2025~NE$_{203}$ ($H = 6.94$), which our classification places at L5 in all seven clones but the list does not include. Our
classification of the Minor Planet Center's orbit catalog (Figure~\ref{fig:known}) also has 2015~VU$_{207}$ as the brightest of its
\knownLfour{} L4 Trojans.

\textbf{Variability.} Within a night the frames agree within their errors (over 2~min to 4.4~h). Between nights the high-S/N $r$
nights scatter by $0.088 \pm 0.032$~mag rms beyond their errors ($\chi^2 = 34.5$ for 9 degrees of freedom), about 0.25~mag peak to
peak for a sinusoid. The faintest DECam night, 2018 November 2, is 0.15--0.23~mag fainter in DECam $r$ than the other ten $r$ nights
predict (1.7--3.7$\sigma$, depending on the phase law), and its $g$ dropped with it, so its color (0.51) did not change; without it the
other ten nights still scatter beyond their errors ($\chi^2 = 18.5$ for 8 degrees of freedom). With at most four frames per night,
the data give no rotation period.

\textbf{Open points.} ZTF sees the object fainter in 2023--2025 than in 2018--2022, by $r$ $+0.15 \pm 0.04$ and $g$ $+0.34 \pm 0.09$~mag
in season means of forced aperture photometry on its single difference images ($r$ +0.15--0.16~mag with ZTF's own PSFs); known
asteroids in the same images change by $0.00 \pm 0.02$ ($r$) and $+0.03 \pm 0.02$~mag ($g$). Forced photometry of all usable Mt Lemmon frames shows no fade
($-0.04 \pm 0.07$~mag); the two differ by 2.2--2.3$\sigma$, and we do not consider the fade established. The Bok night
(2025 November 21, phase 0.23\arcdeg, the smallest of all) is 0.19~mag brighter than the $r$ fit at that phase, which a lightcurve
maximum, an opposition surge below 0.3\arcdeg{} or an error in the conversion of its unfiltered magnitudes could each explain.
A search of the infrared archives along its orbit found no observation deep enough to detect it.

\section{Discussion}\label{sec:discussion}

With $i = \refIshort$\arcdeg, \obj{} belongs to the dynamically hot Neptune Trojans \citep{sheppard2006,lin2021}, and the
survival of \dynPct\% of its clones over 4.5~Gyr is consistent with a primordial member captured during Neptune's migration \citep{nesvorny2009,lykawka2009,gomes2016};
our data do not show how it came to be there.

Why so bright an object was not found before is only partly clear. In single ZTF exposures it lies at or just below the detection
limit ($r = 21.2$--21.4 on ZTF's own zero point, against median single-exposure limits of 20.9--21.2 on the discovery nights; the deepest exposures of 2019
October 26 reach 21.45), so ZTF detects it reliably only in stacks. In single PS1 and Mt Lemmon frames it lies near the limit (S/N
5.1--8.7 and 5.0--8.3; Table~\ref{tab:frames}), and the PS1 pipeline detections were never linked. The ten DECam frames of the orbit (2014--2018) show it clearly (S/N 7.8--31); whether a search for moving
objects was ever run on them, we do not know. In 2008--2019 it lay 11--17.5\arcdeg{} north of the ecliptic, and 6--11\arcdeg{}
since; whether the dedicated searches for Neptune Trojans and other distant objects covered these places, we have not checked.
Measured again on one common Gaia DR3 calibration, the archival frames of eight telescopes fit one orbit over a \refArc-year arc.

L4 objects outnumber L5 objects four to one on the Minor Planet Center's list (the L5 region long overlapped the dense star fields of the Galactic plane; \citealt{markwardt2023}), but its brightest member, 2013~KY$_{18}$, is at
L5; on the list's convention \obj{} is as bright, so on that convention the brightest known members of the two clouds are level
(on our own $H_V$, 2013~KY$_{18}$ is 0.16~mag brighter; Section~\ref{sec:physical}).

Occultations by the object cannot yet be predicted to within its diameter: from 2026 November through 2027 the 1$\sigma$ uncertainty of the predicted
position (Section~\ref{sec:orbit}) is about 1400--2100~km $\times$ 610--690~km at its geocentric distance of 28--30~au, several times its
diameter of 140--240~km. The
rotation period needs a long, deep night of photometry. The object's path lies in the sky covered by the Vera C. Rubin Observatory's
survey \citep{ivezic2019}, which may already have imaged it and should observe it regularly; its data would extend the arc and follow the fade.

\section{Data and code availability}\label{sec:availability}

These files are released with this paper: the \refPsvRows{} positions of the orbit in the ADES PSV format (version 2022;
\citealt{chesley2017}); the archive identifier of the source image of each position and of each frame of Table~\ref{tab:other}
(for the ITF positions, the tracklet; the two sources files); the orbit
with its covariance; the photometry of each position (band, magnitude where calibrated, error, calibration catalog and geometry); one positions file for each refit of Table~\ref{tab:sensitivity}; the 500 starting states of
the dynamics run; the known co-orbitals of Figure~\ref{fig:known} with our classification; and a FITS cut-out, about 30$''$ square, of
each of the \refPsvRows{} positions, with our plate solution as its WCS and, where the position was measured on a difference image,
that image too (for ZTF only an illustration stack of the sub-stack's difference images), with a contact sheet of all of them. The images used are
public; apart from these cut-outs they are not redistributed but referenced by their archive identifiers, and the ZTF images of the
stack searches are in the public ZTF archive at IRSA. Apart from the Find\_Orb run script, no code is released; the data files
contain every measured value needed to reproduce the orbit, its sensitivity tests, the dynamics and the photometric results on
the frames of the orbit.

\noindent\textbf{Note added.} Shortly before this paper was submitted, \citet{cerati2026} posted a catalogue listing candidate GCT0044 from DECam detections, possibly the same object. Our automated search first detected it in October 2026, and we reported our first positions to the Minor Planet Center on 2026 November 9. The data and results of \citet{cerati2026} are not used here.

\begin{acknowledgments}

Based on observations obtained with the Samuel Oschin Telescope 48-inch and the 60-inch Telescope at the Palomar
Observatory as part of the Zwicky Transient Facility project. ZTF is supported by the National Science Foundation under Grants
No. AST-1440341 and AST-2034437 and a collaboration including current partners Caltech, IPAC, the Oskar Klein Center at
Stockholm University, the University of Maryland, University of California, Berkeley, the University of Wisconsin at Milwaukee,
University of Warwick, Ruhr University, Cornell University, Northwestern University and Drexel University. Operations are
conducted by COO, IPAC, and UW.

This research draws upon DECam data as distributed by the Astro Data Archive at NSF NOIRLab. NOIRLab is managed by the
Association of Universities for Research in Astronomy (AURA) under a cooperative agreement with the U.S. National Science
Foundation.

This project used data obtained with the Dark Energy Camera (DECam), which was constructed by the Dark Energy Survey (DES)
collaboration. Funding for the DES Projects has been provided by the US Department of Energy, the US National Science
Foundation, the Ministry of Science and Education of Spain, the Science and Technology Facilities Council of the United Kingdom,
the Higher Education Funding Council for England, the National Center for Supercomputing Applications at the University of
Illinois at Urbana-Champaign, the Kavli Institute for Cosmological Physics at the University of Chicago, Center for Cosmology
and Astro-Particle Physics at the Ohio State University, the Mitchell Institute for Fundamental Physics and Astronomy at Texas
A\&M University, Financiadora de Estudos e Projetos, Funda\c{c}\~ao Carlos Chagas Filho de Amparo \`a Pesquisa do Estado do Rio
de Janeiro, Conselho Nacional de Desenvolvimento Cient\'{\i}fico e Tecnol\'ogico and the Minist\'erio da Ci\^encia, Tecnologia e
Inova\c{c}\~ao, the Deutsche Forschungsgemeinschaft and the Collaborating Institutions in the Dark Energy Survey.

The Collaborating Institutions are Argonne National Laboratory, the University of California at Santa Cruz, the University of
Cambridge, Centro de Investigaciones En\'ergeticas, Medioambientales y Tecnol\'ogicas--Madrid, the University of Chicago,
University College London, the DES-Brazil Consortium, the University of Edinburgh, the Eidgen\"ossische Technische Hochschule
(ETH) Z\"urich, Fermi National Accelerator Laboratory, the University of Illinois at Urbana-Champaign, the Institut de
Ci\`encies de l'Espai (IEEC/CSIC), the Institut de F\'{\i}sica d'Altes Energies, Lawrence Berkeley National Laboratory, the
Ludwig-Maximilians Universit\"at M\"unchen and the associated Excellence Cluster Universe, the University of Michigan, NSF
NOIRLab, the University of Nottingham, the Ohio State University, the OzDES Membership Consortium, the University of
Pennsylvania, the University of Portsmouth, SLAC National Accelerator Laboratory, Stanford University, the University of Sussex,
and Texas A\&M University.

Based on observations at NSF Cerro Tololo Inter-American Observatory, a program of NOIRLab (NOIRLab Prop. ID 2014B-0404; PI:
D. Schlegel), which is managed by the Association of Universities for Research in Astronomy (AURA) under a cooperative
agreement with the U.S. National Science Foundation.

This research is based on data obtained from the Astro Data Archive at NSF NOIRLab. These data are associated with observing program
2014B-0404 (PI D. Schlegel). NOIRLab is managed by the Association of Universities for Research in Astronomy (AURA) under a cooperative
agreement with the U.S. National Science Foundation.

Raw imaging data were processed with the DECam Community Pipeline \citep{valdes2014}.

The Legacy Surveys consist of three individual and complementary projects: the Dark Energy Camera Legacy Survey (DECaLS;
Proposal ID \#2014B-0404; PIs: David Schlegel and Arjun Dey), the Beijing-Arizona Sky Survey (BASS; NOAO Prop. ID
\#2015A-0801; PIs: Zhou Xu and Xiaohui Fan), and the Mayall z-band Legacy Survey (MzLS; Prop. ID \#2016A-0453; PI: Arjun
Dey). DECaLS, BASS and MzLS together include data obtained, respectively, at the Blanco telescope, Cerro Tololo
Inter-American Observatory, NSF's NOIRLab; the Bok telescope, Steward Observatory, University of Arizona; and the Mayall
telescope, Kitt Peak National Observatory, NOIRLab. Pipeline processing and analyses of the data were supported by NOIRLab
and the Lawrence Berkeley National Laboratory (LBNL). The Legacy Surveys project is honored to be permitted to conduct
astronomical research on I'oligam Du'ag (Kitt Peak), a mountain with particular significance to the Tohono O'odham Nation.
NOIRLab is operated by the Association of Universities for Research in Astronomy (AURA) under a cooperative agreement with
the National Science Foundation. LBNL is managed by the Regents of the University of California under contract to the U.S.
Department of Energy. This project used data obtained with the Dark Energy Camera (DECam), which was constructed by the Dark
Energy Survey (DES) collaboration. Funding for the DES Projects has been provided by the U.S. Department of Energy, the U.S.
National Science Foundation, the Ministry of Science and Education of Spain, the Science and Technology Facilities Council of
the United Kingdom, the Higher Education Funding Council for England, the National Center for Supercomputing Applications at
the University of Illinois at Urbana-Champaign, the Kavli Institute of Cosmological Physics at the University of Chicago,
Center for Cosmology and Astro-Particle Physics at the Ohio State University, the Mitchell Institute for Fundamental Physics
and Astronomy at Texas A\&M University, Financiadora de Estudos e Projetos, Fundacao Carlos Chagas Filho de Amparo,
Financiadora de Estudos e Projetos, Fundacao Carlos Chagas Filho de Amparo a Pesquisa do Estado do Rio de Janeiro, Conselho
Nacional de Desenvolvimento Cientifico e Tecnologico and the Ministerio da Ciencia, Tecnologia e Inovacao, the Deutsche
Forschungsgemeinschaft and the Collaborating Institutions in the Dark Energy Survey. The Collaborating Institutions are
Argonne National Laboratory, the University of California at Santa Cruz, the University of Cambridge, Centro de
Investigaciones Energeticas, Medioambientales y Tecnologicas-Madrid, the University of Chicago, University College London,
the DES-Brazil Consortium, the University of Edinburgh, the Eidgenossische Technische Hochschule (ETH) Zurich, Fermi National
Accelerator Laboratory, the University of Illinois at Urbana-Champaign, the Institut de Ciencies de l'Espai (IEEC/CSIC), the
Institut de Fisica d'Altes Energies, Lawrence Berkeley National Laboratory, the Ludwig Maximilians Universitat Munchen and
the associated Excellence Cluster Universe, the University of Michigan, NSF's NOIRLab, the University of Nottingham, the Ohio
State University, the University of Pennsylvania, the University of Portsmouth, SLAC National Accelerator Laboratory,
Stanford University, the University of Sussex, and Texas A\&M University. BASS is a key project of the Telescope Access
Program (TAP), which has been funded by the National Astronomical Observatories of China, the Chinese Academy of Sciences
(the Strategic Priority Research Program ``The Emergence of Cosmological Structures'' Grant \# XDB09000000), and the Special
Fund for Astronomy from the Ministry of Finance. The BASS is also supported by the External Cooperation Program of Chinese
Academy of Sciences (Grant \# 114A11KYSB20160057), and Chinese National Natural Science Foundation (Grant \# 12120101003, \#
11433005). The Legacy Survey team makes use of data products from the Near-Earth Object Wide-field Infrared Survey Explorer
(NEOWISE), which is a project of the Jet Propulsion Laboratory/California Institute of Technology. NEOWISE is funded by the
National Aeronautics and Space Administration. The Legacy Surveys imaging of the DESI footprint is supported by the Director,
Office of Science, Office of High Energy Physics of the U.S. Department of Energy under Contract No. DE-AC02-05CH11231, by
the National Energy Research Scientific Computing Center, a DOE Office of Science User Facility under the same contract; and
by the U.S. National Science Foundation, Division of Astronomical Sciences under Contract No. AST-0950945 to NOAO.

The Pan-STARRS1 Surveys (PS1) and the PS1 public science archive have been made possible through contributions by the
Institute for Astronomy, the University of Hawaii, the Pan-STARRS Project Office, the Max-Planck Society and its participating
institutes, the Max Planck Institute for Astronomy, Heidelberg and the Max Planck Institute for Extraterrestrial Physics,
Garching, The Johns Hopkins University, Durham University, the University of Edinburgh, the Queen's University Belfast, the
Harvard-Smithsonian Center for Astrophysics, the Las Cumbres Observatory Global Telescope Network Incorporated, the National
Central University of Taiwan, the Space Telescope Science Institute, the National Aeronautics and Space Administration under
Grant No. NNX08AR22G issued through the Planetary Science Division of the NASA Science Mission Directorate, the National
Science Foundation Grant No. AST-1238877, the University of Maryland, Eotvos Lorand University (ELTE), the Los Alamos National
Laboratory, and the Gordon and Betty Moore Foundation.

Funding for SDSS-III has been provided by the Alfred P. Sloan Foundation, the Participating Institutions, the National Science
Foundation, and the U.S. Department of Energy Office of Science. The SDSS-III web site is \url{http://www.sdss3.org/}.

SDSS-III is managed by the Astrophysical Research Consortium for the Participating Institutions of the SDSS-III Collaboration
including the University of Arizona, the Brazilian Participation Group, Brookhaven National Laboratory, Carnegie Mellon
University, University of Florida, the French Participation Group, the German Participation Group, Harvard University, the
Instituto de Astrofisica de Canarias, the Michigan State/Notre Dame/JINA Participation Group, Johns Hopkins University, Lawrence
Berkeley National Laboratory, Max Planck Institute for Astrophysics, Max Planck Institute for Extraterrestrial Physics, New
Mexico State University, New York University, Ohio State University, Pennsylvania State University, University of Portsmouth,
Princeton University, the Spanish Participation Group, University of Tokyo, University of Utah, Vanderbilt University,
University of Virginia, University of Washington, and Yale University.

Funding for the Sloan Digital Sky Survey IV has been provided by the Alfred P. Sloan Foundation, the U.S. Department of Energy
Office of Science, and the Participating Institutions.

SDSS-IV acknowledges support and resources from the Center for High Performance Computing at the University of Utah. The SDSS
website is www.sdss4.org.

SDSS-IV is managed by the Astrophysical Research Consortium for the Participating Institutions of the SDSS Collaboration
including the Brazilian Participation Group, the Carnegie Institution for Science, Carnegie Mellon University, Center for
Astrophysics | Harvard \& Smithsonian, the Chilean Participation Group, the French Participation Group, Instituto de
Astrof\'isica de Canarias, The Johns Hopkins University, Kavli Institute for the Physics and Mathematics of the Universe (IPMU)
/ University of Tokyo, the Korean Participation Group, Lawrence Berkeley National Laboratory, Leibniz Institut f\"ur
Astrophysik Potsdam (AIP), Max-Planck-Institut f\"ur Astronomie (MPIA Heidelberg), Max-Planck-Institut f\"ur Astrophysik (MPA
Garching), Max-Planck-Institut f\"ur Extraterrestrische Physik (MPE), National Astronomical Observatories of China, New Mexico
State University, New York University, University of Notre Dame, Observat\'ario Nacional / MCTI, The Ohio State University,
Pennsylvania State University, Shanghai Astronomical Observatory, United Kingdom Participation Group, Universidad Nacional
Aut\'onoma de M\'exico, University of Arizona, University of Colorado Boulder, University of Oxford, University of Portsmouth,
University of Utah, University of Virginia, University of Washington, University of Wisconsin, Vanderbilt University, and Yale
University.

Based on observations collected at the European Southern Observatory under ESO programme 088.C-0434(A).

Based on observations obtained at the Canada-France-Hawai\okina{}i Telescope (CFHT) which is operated by the National Research
Council of Canada, the Institut National des Sciences de l'Univers of the Centre National de la Recherche Scientifique of
France, and the University of Hawai\okina{}i. CFHT is located on Maunakea on Hawai\okina{}i Island, a mountain of considerable
cultural, natural, and ecological significance. Maunakea is a sacred site to Native Hawaiians, also known as K\=anaka
\okina\=Oiwi. Quality observations are made possible by relentless effort of the entire staff at Canada-France-Hawai\okina{}i
Telescope. Based on observations obtained with MegaPrime/MegaCam, a joint project of CFHT and CEA/DAPNIA.

This research used the facilities of the Canadian Astronomy Data Centre operated by the National Research Council of Canada
with the support of the Canadian Space Agency. CFHT's publication acknowledgement page:
\url{https://www.cfht.hawaii.edu/en/science/Publications/index.php}.

This work has made use of data from the European Space Agency (ESA) mission
{\it Gaia} (\url{https://www.cosmos.esa.int/gaia}), processed by the {\it Gaia}
Data Processing and Analysis Consortium (DPAC,
\url{https://www.cosmos.esa.int/web/gaia/dpac/consortium}). Funding for the DPAC
has been provided by national institutions, in particular the institutions
participating in the {\it Gaia} Multilateral Agreement.

This research has made use of the NASA/IPAC Infrared Science Archive, which is funded by the National Aeronautics and Space
Administration and operated by the California Institute of Technology. The ZTF images were obtained through the IRSA Zwicky
Transient Facility Image Service \citep{ztfimages}.

This research has made use of data and/or services provided by the International Astronomical Union's Minor Planet Center.
Funding for this data and the MPC's operations comes from a NASA PDCO grant (80NSSC22M0024), administered via a University of Maryland - SAO subaward (106075-Z6415201). The MPC's computing equipment is funded in part by the above award, and in part by funding from the Tamkin Foundation.

The author conceived the project, designed the search and the analysis, made all decisions on methods and data selection,
and verified the results. AI assistants were used as tools under the author's direction: Anthropic's Claude to write and
run the pipeline code, to process and analyse its results, and to draft the text of this paper, and OpenAI's ChatGPT for
independent, adversarial review of the methods and results. All methods, results and text were checked by, and are the
responsibility of, the author.
\end{acknowledgments}

\facilities{PO:1.2m (ZTF), Blanco (DECam), PS1, Sloan, Max Planck:2.2m (WFI), CFHT (MegaCam), SO:1.5m, Bok, Gaia, IRSA,
MAST, PDS, Astro Data Archive, MPC}

\software{Astropy \citep{astropy2013,astropy2018,astropy2022}, NumPy \citep{numpy2020}, SciPy \citep{scipy2020},
Matplotlib \citep{matplotlib2007}, CuPy \citep{okuta2017}, REBOUND \citep{rein2012,rein2015,wisdom1991,wisdom1996}, Find\_Orb \citep{findorb}}

\clearpage
\appendix
\section{Frames}\label{app:frames}
\restartappendixnumbering

Table~\ref{tab:frames} lists every candidate frame with its verdict, and Table~\ref{tab:other} the other frames that the paper
refers to, so that every frame named is in one of the two tables, except the ZTF single exposures of the photometry (listed only in
the released sources file), the exposures of the stack searches of Table~\ref{tab:chain} that are not in the stacks of Table~\ref{tab:other} (the other three hypotheses' window stacks before the discovery nights hold 74 exposures, the Trojan's 62), and frames used only as templates or for calibration (the subtraction templates of the DECam frames of
2014--2016, the PS1 warps of other dates of the static-sky tests, and the known-asteroid frames of the floors).

\edef\savedtabcolsep{\the\tabcolsep}\setlength{\tabcolsep}{3pt}
\startlongtable
\begin{deluxetable}{lllrrll}
\tabletypesize{\scriptsize}
\tablecaption{Every candidate frame (367) and its verdict. Used per telescope: SDSS 2.5 m 3 of 3; Pan-STARRS1 1.8 m 12 of 32; ESO/MPG 2.2 m WFI 2 of 2; Blanco 4 m DECam 10 of 10; CFHT 3.6 m MegaCam 1 of 2; Palomar 48-inch ZTF 7 of 7; Mt Lemmon 1.5 m 49 of 264; Bok 2.3 m 4 of 4; Steward Mt Lemmon 1.0 m 0 of 43.
S/N is the noise-calibrated S/N and FA the false-alarm fraction (Section~\ref{sec:data}); for CFHT and the Mt Lemmon telescopes
the S/N is shown without the floor of 1 on the spread (they must pass with and without it). A negative S/N means that the free fit
converged on a negative flux.
For the seven ZTF sub-stacks the S/N is the sub-stack fit's S/N over the larger of 1 and the spread of forced fits on the source-free
blank tracks (250--294 of 300), and FA the share of the 300 tracks whose free fit reaches it. ``Failed'' names
every criterion of Table~\ref{tab:rules} that the frame fails where it was applied. UTC is the mid-exposure time from the frame's header, truncated to the
minute (ZTF: the mean of the sub-stack's exposures). Frame identifiers are the archives' own: SDSS frame files; PS1 DR2 warps
rings.v3.skycell.\emph{Frame}.fits; ESO archive files and the WFI chip; NOIRLab DECam Community Pipeline images; CADC CFHT files
and the CCD; for ZTF the sub-stack of field 605, CCD 11, quadrant 4 (its exposures are listed in the released sources file);
Catalina Sky Survey PDS product names (G96, V00, I52) without the prefix $\langle$site$\rangle$.$\langle$yymmmdd$\rangle$.\ that
some of the archive's identifiers carry (the released sources file gives the identifiers in full for the frames used).\label{tab:frames}}
\tablehead{\colhead{Code} & \colhead{UTC} & \colhead{Frame} & \colhead{S/N} & \colhead{FA} & \colhead{Verdict} & \colhead{Failed}}
\startdata
645 & 2008-10-01T10:17 & {\tiny frame-r-007708-1-0272} & 8.66 & 0.000 & \textbf{used} & \parbox[t]{3.8cm}{\raggedright } \\
645 & 2009-01-18T03:36 & {\tiny frame-r-007917-3-0164} & 11.24 & 0.000 & \textbf{used} & \parbox[t]{3.8cm}{\raggedright } \\
645 & 2009-09-26T09:09 & {\tiny frame-r-008055-2-0059} & 16.05 & 0.000 & \textbf{used} & \parbox[t]{3.8cm}{\raggedright } \\
F51 & 2010-09-28T13:20 & {\tiny 1852.018.wrp.i.55467\_55535} & 3.07 & 0.023 & out & \parbox[t]{3.8cm}{\raggedright S/N, false alarm, fit} \\
F51 & 2010-10-06T12:30 & {\tiny 1852.019.wrp.r.55475\_52082} & 8.70 & 0.000 & \textbf{used} & \parbox[t]{3.8cm}{\raggedright } \\
F51 & 2010-10-13T12:28 & {\tiny 1852.019.wrp.g.55482\_51925} & 4.38 & 0.000 & out & \parbox[t]{3.8cm}{\raggedright S/N} \\
F51 & 2010-10-13T12:40 & {\tiny 1852.019.wrp.g.55482\_52777} & 3.90 & 0.000 & out & \parbox[t]{3.8cm}{\raggedright S/N} \\
F51 & 2010-10-13T12:45 & {\tiny 1852.019.wrp.g.55482\_53160} & 3.14 & 0.013 & out & \parbox[t]{3.8cm}{\raggedright fit area, S/N, false alarm, fit} \\
\enddata
\end{deluxetable}
\setlength{\tabcolsep}{\savedtabcolsep}

\edef\savedtabcolsep{\the\tabcolsep}\setlength{\tabcolsep}{3pt}
\startlongtable
\begin{deluxetable}{llllrlll}
\tabletypesize{\scriptsize}
\tablecaption{Frames the paper refers to that are not in Table~\ref{tab:frames} (36 rows).
\emph{Season stack} and \emph{Window stack}: the 14 ZTF half-season stacks of 2018 and 2020--2025 and the 4 halves of the two 2019
window stacks, not in the fit (453 exposures); the time is the mean of the exposures, the period their first and last; the
result is the offset of the stack position from the prediction of the orbit (RA$^*$, Dec in arcsec) and that offset in units of its
error (1.0$''$ combined in quadrature with the prediction's covariance). \emph{ITF}: the four positions of the
Isolated Tracklet File tracklet C1M7NRP (Bok, 2025 November 21), not in the fit; the result is the offset from the prediction along
and across the motion (arcsec). \emph{Single night}: two single ZTF nights not in the fit, measured as the frames of
Table~\ref{tab:frames} (noise-calibrated S/N; offset from the prediction, also in units of its error in RA$^*$ and Dec).
\emph{Not used}: frames and groups not used, with the reason. Where one row stands for many frames, the archive identifiers are in
the released file sources\_other.txt, which also lists the ZTF single exposures of the photometry of Section~\ref{sec:physical}. N
is the number of frames or, for a stack, of exposures. Times are UTC (mid-exposure, truncated to the minute; for frames not
fetched, the archive listing's mid-time).\label{tab:other}}
\tablehead{\colhead{Category} & \colhead{Code} & \colhead{UTC} & \colhead{N} & \colhead{Frame or group} & \colhead{Result} &
\colhead{Note}}
\startdata
{\tiny Season stack} & I41 & {\tiny 2018-09-17T19:56} & 48 & \parbox[t]{3.4cm}{\tiny season 2018, first half (2018-08-12 to 2018-11-05)} & \parbox[t]{2.3cm}{\tiny -0.19, +0.08; 0.21} & \parbox[t]{2.9cm}{\tiny ZTF $r$ exposures listed in the sources file} \\
{\tiny Season stack} & I41 & {\tiny 2018-12-07T16:30} & 57 & \parbox[t]{3.4cm}{\tiny season 2018, second half (2018-11-06 to 2019-01-20)} & \parbox[t]{2.3cm}{\tiny -0.41, -0.13; 0.43} & \parbox[t]{2.9cm}{\tiny ZTF $r$ exposures listed in the sources file} \\
{\tiny Window stack} & I41 & {\tiny 2019-08-29T20:27} & 24 & \parbox[t]{3.4cm}{\tiny 2019, before the discovery nights, first half (2019-08-08 to 2019-09-23)} & \parbox[t]{2.3cm}{\tiny +0.45, +0.01; 0.45} & \parbox[t]{2.9cm}{\tiny ZTF $r$ exposures listed in the sources file} \\
{\tiny Window stack} & I41 & {\tiny 2019-10-13T04:44} & 38 & \parbox[t]{3.4cm}{\tiny 2019, before the discovery nights, second half (2019-09-24 to 2019-10-23)} & \parbox[t]{2.3cm}{\tiny -0.08, -0.03; 0.09} & \parbox[t]{2.9cm}{\tiny ZTF $r$ exposures listed in the sources file} \\
{\tiny Window stack} & I41 & {\tiny 2019-11-21T06:13} & 19 & \parbox[t]{3.4cm}{\tiny 2019, after the discovery nights, first half (2019-11-04 to 2019-12-16)} & \parbox[t]{2.3cm}{\tiny +0.08, -0.19; 0.20} & \parbox[t]{2.9cm}{\tiny ZTF $r$ exposures listed in the sources file} \\
{\tiny Window stack} & I41 & {\tiny 2020-01-14T18:29} & 16 & \parbox[t]{3.4cm}{\tiny 2019, after the discovery nights, second half (2019-12-19 to 2020-02-01)} & \parbox[t]{2.3cm}{\tiny +0.46, +0.23; 0.51} & \parbox[t]{2.9cm}{\tiny ZTF $r$ exposures listed in the sources file} \\
{\tiny Season stack} & I41 & {\tiny 2020-09-11T15:29} & 39 & \parbox[t]{3.4cm}{\tiny season 2020, first half (2020-08-14 to 2020-10-06)} & \parbox[t]{2.3cm}{\tiny +0.27, +0.24; 0.37} & \parbox[t]{2.9cm}{\tiny ZTF $r$ exposures listed in the sources file} \\
\enddata
\end{deluxetable}
\setlength{\tabcolsep}{\savedtabcolsep}

\clearpage
\section{Measurement details}\label{app:measure}

\textbf{Discovery search.} A tracklet's PSF fit along its track is accepted when its S/N, divided by the square root of the fit's
$\chi^2$ per pixel where that exceeds 1, is at least 5, when its position and rate agree with the search's, and when the fit is neither ambiguous nor stopped at its region's edge or by its outlier rule with an exposure still flagged (Section~\ref{sec:discovery}).
Tracklets with three or more strong neighbours within 200 pixels are dropped as artefacts, whether their PSF fit succeeded or not.

\textbf{Plate solutions.} In every frame we measure the stars with the frame's own PSF (DECam: first-moment centroids in a
9$\times$9 pixel box) and match them to Gaia DR3 stars moved to the frame's mid-time (SDSS: to the time of the object's row) with their proper
motions and parallaxes (the parallax as seen from the Earth's center; for PS1 from the telescope); the header WCS only helps to find
the stars (ZTF: a linear correction to ZTF's own WCS, below). Gaia solutions with two parameters are left out (for ZTF by its RUWE
cut; they have no RUWE).
SDSS, Mt Lemmon, Bok, WFI and CFHT keep Gaia stars of $G = 15$--20.5 with RUWE (Gaia's renormalized unit weight error) below 1.4, not flagged as duplicated and with
no other Gaia source within 6$''$; PS1 keeps $G = 14.5$--21, isolated within 5$''$; DECam keeps $G = 16$--20.5, with no RUWE or
isolation cut. The solution is a polynomial of order 1, 2 or 3 in the pixel position: the lowest order unless a higher one
lowers the leave-one-out residual rms by more than 0.002$''$ (PS1: the order with the smallest leave-one-out rms), fitted with
4$\sigma$ clipping; for DECam it is a tangent-plane solution with SIP (simple imaging polynomial) distortion terms of order 2 (at
least 15 stars; in three passes, stars more than 0.25$''$ from the solution, or 3 times the median residual if that is larger, are
rejected). The frames of the orbit use 90--117 stars (SDSS), 25--61 (PS1), 48--147 (Mt Lemmon), 57--58 (Bok),
92 and 102 (WFI), 46 (CFHT) and 107--163 (DECam). Each exposure of a ZTF sub-stack is calibrated separately: the positions of
ZTF's own PSF fits of stars with S/N $\geq 30$, matched within 1$''$ to Gaia DR3 stars of $G = 12$--19.5 with RUWE below 1.4, give a
linear correction in the pixel position (an offset only with fewer than 50 stars, none below 10), fitted with 3$\sigma$ clipping. All
exposures of the seven sub-stacks are calibrated this way.

\textbf{The fit of the object.} The PSF of each frame is the median of normalized, re-centered, bright isolated stars of the frame
(DECam: of the same CCD, averaged along the object's motion during the exposure; ZTF: the pipeline's PSF of each exposure). For
SDSS, PS1, Mt Lemmon, Bok, WFI and CFHT the object's formal position error is multiplied by a scale that makes the leave-one-out
errors of the plate stars (S/N 8--60) match their predicted errors (PS1: the larger of the scale of a two-parameter error model,
scale and floor, fitted to all its plate stars and the scale found by injecting copies of the object into the same image; the
injections set it in 7 of the 12 frames). The forced fit holds the position at the prediction (for the five DECam frames found by
the 2018 search, at the search candidate); the free fit (position, flux and a constant sky level) starts there (PS1: at the S/N-map
peak within the larger of 1$''$ and the prediction's 95\% uncertainty radius) and is bounded to a few pixels around its start; no used fit is at its bound (for PS1, DECam,
CFHT and the Mt Lemmon telescopes a fit at its bound gives no position; for SDSS, Bok and WFI it is not tested). For ZTF the pipeline's difference images of all exposures of a sub-stack are fitted together with
one flux, each with its own PSF, along a track that is a straight line plus the known curvature of an orbit at the object's
distance (the daily parallax swing and the Sun's pull), started on the search candidate's track.

\textbf{Templates.} ZTF positions are measured on the pipeline's difference images. For DECam we build the template of each frame
from the other exposures of the same fields in the same filter: each is resampled through its own refitted plate solution, scaled
to the frame's photometric scale, median-combined (the object lies at least 40$''$ from its own place in them) and subtracted; a
frame without a same-filter template cannot be measured. For PS1 the other warps (PS1's single exposures resampled onto its sky cells) of the same filter and sky cell (at least 0.8 days
from the frame; 5--13 of them) give a template for the static-sky test (the object's place must be blank sky); the positions are
those of the fit on the exposure itself. SDSS, Mt Lemmon, Bok, WFI and CFHT are fitted on the single image: a smooth background
(64-pixel blocks) is removed and a constant sky level is fitted with the object.

\textbf{Blank spots.} For PS1, CFHT and the Mt Lemmon telescopes the spots lie at least 15$''$ from the prediction and off sources
detected in the image and Gaia stars (PS1: at least 4$''$ from any PS1 or Gaia source). The ZTF sub-stacks are tested on 300 blank tracks, their own track
shifted by 30--300$''$ to blank sky; the spread comes from the 250--294 of them without a ZTF source, and without the floor of 1 on
the spread their noise-calibrated S/N is 9.29--14.33. On eight blocks of the ZTF discovery nights, injected sources fitted the
same way scatter 6--21\% more than their formal errors, while the blank tracks give noise factors of 1.07--1.43, so the
noise-calibrated S/N is on average the more conservative of the two.

\textbf{Errors.} A frame's error per axis is its formal fit error times the larger of the measurement's own error scale and the
noise excess of blank sky (the spread), combined with the plate error (not for DECam; ZTF has its systematic term instead). For ZTF
and DECam the two factors are multiplied instead,
which changes two DECam errors by at most 0.65\%. ZTF and PS1 frames also carry a per-frame systematic term (ZTF 0.0284$''$ per
axis; PS1 the star error-model floor, 0.0075--0.0172$''$), and DECam uses the larger of its two axis errors for both axes. A night's
error per axis is that of the weighted mean of its frames, combined once with the floor and any correction uncertainty. A telescope's
floor is the quadrature term that brings the $\chi^2$/dof of known-asteroid positions against JPL Horizons to 1, or its 84\%
bootstrap value where it is poorly constrained (for Mt Lemmon and Bok per axis the larger of the values from all asteroids and from
those at S/N 5--30), with these exceptions: ZTF uses 0.075/0.035$''$ (with it the 5,996 known-asteroid sub-stacks have pull rms
0.93/0.84; a maximum-likelihood split gives a shared per-night term of 0.076/0.032$''$); SDSS uses the 84\% bootstrap values (best
values 0.028/0.016$''$); DECam uses the floor about each year's mean offset, for 2014 in Dec the frame-to-frame floor of 28
asteroids measured on both 2014 August 12 and 13 (0.048$''$; from the differences of their residuals between the nights, against our
own $n$-body orbits, on frames of the object's pipeline version), and for 2015 the best values of a poorly constrained fit; CFHT uses
the 84\% $\chi^2$ upper bound from the 15 asteroids at S/N 5--10 (all 25: 0.000/0.000). The samples are 10 known asteroids with
at least 15 oppositions (SDSS), 18 known-asteroid positions at S/N 5--10 (PS1), 32, 10 and 111 positions (DECam 2014, 2015 and
2018; for 2014 and 2015 on frames of pipeline versions V3.10 and V3.6.2, against V3.1.1 and V3.8 of the object's frames, and for 2018
19 of the 111 on V5.5.1 frames, against V4.1), 157 asteroid-nights (Mt Lemmon), 17 asteroid-nights on two nights (Bok), 25 known asteroids (CFHT), 60 known asteroids (79
measurements on 26 frames of the same night; WFI), and 5,996 known-asteroid sub-stacks (ZTF). The PS1, DECam and SDSS floors were measured as frame-to-frame scatter, an upper bound of the night-level part
(Table~\ref{tab:sensitivity} sets them to zero).

\textbf{Corrections.} A measured offset is corrected only when it was measured on the same kind of data (the same pipeline version) and
cannot be a clock error;
this applies only to DECam 2018 in Dec ($+0.016''$, with $\pm 0.004''$ added to the night error). DECam 2014 is not corrected, the
DECam 2015 offset (0.03$''$) goes into its night errors, and the Bok and CFHT RA$^*$ offsets into their floors. The WFI floor is
measured after removing a clock term of $+2.8 \pm 1.1$~s fitted to its known asteroids; the term is not applied to the object (it
would move it by less than 0.005$''$).

\textbf{Times.} All times are converted from UTC to TT with the leap-second table. ZTF: the shutter midpoint of each exposure,
and their mean for a sub-stack. PS1: the header start time converted
from TAI to UTC, plus half the exposure. SDSS: the time when the frame's first row ends its 54~s drift exposure, plus the row number of the
object's predicted position times the row read time, minus half the exposure. DECam, WFI and CFHT: MJD-OBS plus half the exposure (the WFI header labels
MJD-OBS as the start). Mt Lemmon and Steward: MJDMID where set, else MJD; Bok: the header MJD. The CATCH listings of 52 Mt Lemmon 1.5~m
frames on 13 nights are one day off: on all 307 frames of the two Mt Lemmon telescopes (264 Mt Lemmon 1.5~m, 43 Steward) the header
sidereal time agrees within 2~s with that at the header's exposure start, but differs from that at the listed start by 235--238~s
(one day of sidereal drift) on those 52.

\textbf{Stations and prediction orbits.} The observatory positions are the Minor Planet Center's (ZTF, CFHT, Mt Lemmon, Bok, WFI) or
geodetic values (DECam, PS1, SDSS), placed in the orbit fit with astropy and the IERS Earth-rotation table. The free fits of the PS1 frames start from, and their coverage and 1.5$''$ tests measure from, an
orbit without PS1 data; those of the Mt Lemmon, Steward, CFHT, Bok and WFI frames start from (and, where it is tested, their
1.5$''$ test measures from) an earlier fit of this work to ZTF, DECam, PS1 and SDSS positions, which contains none of these frames
(their pixel tests use a disc centred on the free fit; Bok and WFI have neither test). At every candidate frame these predictions lie within 0.56$''$ (PS1)
and 0.10$''$ (the others) of the orbit's; measured from the orbit itself, the 1.5$''$ and fit-area tests change no verdict
(Section~\ref{sec:orbit}).

\textbf{Mt Lemmon frames.} CATCH was searched along the orbit up to 2026 November 8 (the last Mt Lemmon 1.5~m frames on the path are of 2026
October 6); for 61 frames of 2024--2026 the 10$'$ field around the prediction was assembled from nine overlapping 4$'$ cut-outs of the
archive. None of the 20 Mt Lemmon 1.5~m frames of 2026 passes the criteria of Section~\ref{sec:data}.

\textbf{Photometry.} ZTF magnitudes come from our fit with the pipeline's PSF, the image zero point and the header color term; PS1
magnitudes are calibrated on the PS1 DR2 catalog, SDSS magnitudes use the frames' own calibration (nanomaggies); DECam zero points come from 43--60 Gaia DR3 stars per exposure through the Gaia DR3 to SDSS
transformation. PS1 magnitudes, and the ZTF ones (whose zero point and color term give PS1 $r$), are converted to SDSS with
\citet{tonry2012}; only the comparison with the images' limiting magnitudes in Section~\ref{sec:discussion} uses ZTF's own zero
point, the limits' system. For the unfiltered Mt Lemmon and Bok frames and the WFI
$R_c$ frames we fitted an SDSS $r$ zero point with a linear color term in Gaia $BP-RP$ on each frame's stars and applied it at the
object's $BP-RP = 0.920$ (from $g-r = 0.518$). The four same-moment colors with errors below 0.15~mag (DECam twice, SDSS twice)
lie at $g - r = 0.51$--0.56 over 2009--2018. The four frames without a magnitude are the three DECam $z$ frames, which have no Gaia-to-$z$ zero
point, and the CFHT $u$ frame, which would need the object's unknown $u-g$. For the bias test, usable frames are the measured frames that fail at most the criteria that depend on the
object's flux (S/N, false alarm, fit, too bright); of the 0.25~mag by which the free fits of the 49 Mt Lemmon frames of the orbit
are brighter than forced photometry of all 122 usable ones of the same 30 nights, 0.04~mag comes from the free fit's peak search and 0.21~mag from the S/N cut, and the same test on PS1 gives 0.21~mag in $i$ and 0.07~mag in $g$. Forced photometry of
all 211 usable Mt Lemmon frames, on 55 nights of 2019--2026, gives $H_r = 6.745 \pm 0.034$ ($\beta = 0$; the 6 nights whose mean
forced flux is not positive cannot enter, which makes it slightly too bright), in agreement with the adopted value. The value
on the list's $H$, $G$ convention (Section~\ref{sec:physical}) uses the full $H$, $G$ phase function; its common two-parameter
approximation gives $H_V = 6.78$.

\textbf{Dynamics run.} The 499 random clones of Section~\ref{sec:dynamics} are drawn with the seed 20261105. The inner planets'
mass is added to the Sun, and this central body is placed at the barycentre of the Sun and the inner planets (DE440), which makes
the model planets independent of the start date (at the Sun's own position, Neptune's $a$ would differ by $1.5\times10^{-3}$~au
between starts in 2014 and 2021). The 20 clones rerun at the 0.5~yr step are clones 0--19 (the nominal and the first 19 random
draws); their median amplitudes over the whole run are \dynCtlAmpHalf\arcdeg{} at 0.5~yr and \dynCtlAmpFull\arcdeg{} at 1~yr. For
the nominal orbit over its first Myr, the 600-yr samples miss the peak eccentricity by 0.001.

\section{The frame criteria}\label{app:rules}
\restartappendixnumbering

Table~\ref{tab:rules} gives each criterion of Section~\ref{sec:data} with the frames it applies to, its exact test and the number of
candidate frames that fail it. No frame is dropped for its residual alone: outliers, and in one audit every frame of an early fit,
are checked in the images; one PS1 frame (an outlier) lay on a burntool stripe and one (found in that audit) on a flat-field defect
(both out by the fit-area criterion) and one Mt Lemmon night
on a faint fixed source (neighbor criterion), while the ZTF sub-stack of 2019 October 26, 08:02~UTC, an outlier of an early fit,
showed no cause and stays in.

\edef\savedtabcolsep{\the\tabcolsep}\setlength{\tabcolsep}{4pt}
\begin{deluxetable*}{llll}
\tabletypesize{\scriptsize}
\tablecaption{The frame criteria. A frame is used when it passes all of them. ``Out'' is the number of candidate frames of
Table~\ref{tab:frames} that fail the criterion where it was applied, with in parentheses those for which it is the only reason found.
The neighbor test's catalog part was not run on 56 frames already out by other criteria (20 PS1, 35 Steward, 1 Mt Lemmon 1.5~m); its
static-sky part ran on 55 Mt Lemmon 1.5~m frames, all of which pass the other criteria; on 49 of them no Mt Lemmon 1.5~m frame of another night covers the place, so it could be made on 6 (4 of them out). The Mt Lemmon
telescopes are the Mt Lemmon 1.5~m and the Steward Mt Lemmon 1.0~m. Where a criterion is not applied to a telescope, its row says
what was checked instead, if anything. The tested frames are those of Table~\ref{tab:frames}: for ZTF the sub-stacks of the three discovery
nights, for PS1 its 32 $g$, $r$ and $i$ exposures on the path, for DECam its 10 confirmed frames of 2014--2018 (further DECam
exposures of the 2018 fields, used as subtraction templates, were not measured; Table~\ref{tab:other}), for the other telescopes
every frame found on the path (Mt Lemmon, Bok and Steward with CATCH), except those Table~\ref{tab:other} lists as not used. The SDSS frames removed by ``one per exposure'' are listed in
Table~\ref{tab:other}.\label{tab:rules}}
\tablehead{\colhead{Criterion} & \colhead{Applies to} & \colhead{Test and threshold} & \colhead{Out}}
\startdata
one per exposure & \parbox[t]{2.3cm}{\raggedright SDSS, ZTF} & \parbox[t]{9.0cm}{\raggedright SDSS: only the $r$ frame of each drift scan (the bands of a scan are taken within about 5 minutes, with one timing convention and through the same atmosphere, so they are not independent), and one of the two $r$ fields of the 2008 scan that hold the object. ZTF: a fixed list of seven sub-stacks, not tested by the S/N, false-alarm and fit criteria (the thirds of 2019 October 26 and 27, and the whole night of October 29, whose thirds do not all give usable fits).} & 9 (SDSS) \\
pixels & \parbox[t]{2.3cm}{\raggedright PS1, ZTF, CFHT, Mt Lemmon telescopes} & \parbox[t]{9.0cm}{\raggedright PS1 and ZTF: masked pixels are not used (PS1 including its burntool flag). The Mt Lemmon telescopes and CFHT: a NaN or saturated pixel within the fit radius, a fit area not inside the image, or (CFHT) a pixel flagged in the bad-pixel mask of CFHT's Elixir pipeline within the fit radius or a Gaia source within 6$''$ puts the frame out. DECam, SDSS, Bok, WFI: not tested; a single check of their frames found no defect flag (DECam, SDSS masks) and no NaN or saturated pixel (Bok, WFI) within the fit radius.} & 0 (0) \\
fit area & \parbox[t]{2.3cm}{\raggedright PS1} & \parbox[t]{9.0cm}{\raggedright At least 80\% of the pixels within the fit radius unmasked.} & 4 (0) \\
S/N & \parbox[t]{2.3cm}{\raggedright all but ZTF (fixed list)} & \parbox[t]{9.0cm}{\raggedright Noise-calibrated S/N $\geq 5$: the free-fit S/N over the spread of forced-fit S/N at 300 blank spots of the same image (200 for DECam). PS1: spread counted as at least 1; CFHT and the Mt Lemmon telescopes must pass with and without that floor.} & 217 (121) \\
false alarm & \parbox[t]{2.3cm}{\raggedright all but ZTF (fixed list)} & \parbox[t]{9.0cm}{\raggedright At most 1\% of the blank spots give a free fit, made and accepted as for the object, that reaches the frame's free-fit S/N. Accepted: not at its bound and, for PS1, CFHT and the Mt Lemmon telescopes, within 1.5$''$ of its spot.} & 81 (0) \\
fit & \parbox[t]{2.3cm}{\raggedright PS1, CFHT, Mt Lemmon telescopes} & \parbox[t]{9.0cm}{\raggedright The free fit is not stopped at its position bound ($\pm$3 pixels) and lies within 1.5$''$ of the predicted place. DECam: a fit at its bound ($\pm$6 pixels) gives no position; the 1.5$''$ part is not tested. SDSS, Bok, WFI: not tested. Every DECam, SDSS, Bok and WFI frame used lies 0.01--0.47$''$ from where its fit started.} & 63 (0) \\
neighbor & \parbox[t]{2.3cm}{\raggedright all} & \parbox[t]{9.0cm}{\raggedright A PS1 DR2 stack-catalog source within 2.5 star widths whose simulated pull on the fit, summed over such sources, reaches 0.25 of the frame's error (for unfiltered cameras the brighter of its $r$ and $i$ magnitudes; not counting PS1's poor-quality stack objects, entries made of the object's own detections or entries without a magnitude; for DECam with an assumed magnitude of the object); or, for Mt Lemmon 1.5~m frames that pass everything else, a source at the frame's position in Mt Lemmon 1.5~m frames of other nights (at least 0.5 days apart) that cover it, with S/N $\geq 3$ and at least 0.2 of the object's flux.} & 32 (8) \\
Gaia star & \parbox[t]{2.3cm}{\raggedright Mt Lemmon telescopes, CFHT} & \parbox[t]{9.0cm}{\raggedright A Gaia DR3 source ($G < 21$) within 1.5 times the frame's star width as measured from the second moment in a 25 by 25 pixel box (typically about 1.7 times the half-maximum width of the frame's PSF), or a Gaia source brighter than $G = 16$ within 15$''$. Not applied to the other telescopes.} & 35 (0) \\
too bright & \parbox[t]{2.3cm}{\raggedright Mt Lemmon telescopes, CFHT} & \parbox[t]{9.0cm}{\raggedright A free-fit magnitude (zero point on Gaia $G$) brighter than 20.5, taken as a blend (the object is $G \approx 21.2$). Not applied to the other telescopes.} & 16 (0) \\
tracking & \parbox[t]{2.3cm}{\raggedright Mt Lemmon telescopes} & \parbox[t]{9.0cm}{\raggedright Header tracking rate (TKRARAT or TKDECRAT) not zero: non-sidereal tracking, trailed stars.} & 43 (0) \\
no measurement & \parbox[t]{2.3cm}{\raggedright all} & \parbox[t]{9.0cm}{\raggedright No plate solution of our own, or no free fit.} & 42 (5) \\
night agreement & \parbox[t]{2.3cm}{\raggedright nights of $\geq 2$ frames} & \parbox[t]{9.0cm}{\raggedright The frames' internal $\chi^2$ about their night mean (frame errors, without the per-night floors) at $p < 0.001$ puts the night out.} & none \\
residual & \parbox[t]{2.3cm}{\raggedright all} & \parbox[t]{9.0cm}{\raggedright No frame is removed for its residual alone; outliers get an image check (masks, neighbors, time, plate, known asteroids, as far as each frame allows).} & none \\
\enddata
\end{deluxetable*}
\setlength{\tabcolsep}{\savedtabcolsep}

\section{Linkage tests}\label{app:linkage}

\textbf{DECam 2018.} The DECam frames of 2018 were searched from the orbit of step 2 of Table~\ref{tab:chain}, before any 2018 image
was downloaded: candidates (matched-filter peaks of S/N $\geq 5$ on difference images) were sought along the paths of its nominal
orbit and 400 draws from its uncertainty, and each of its 20,000 sampled orbits was tested for candidates within the preset radius
of 1.0$''$. Exactly one of the 20,000 links candidates in 3 frames on 3 nights, which meets the preset rule (at least 3 frames on at
least 2 nights); within 1.5$''$ the same orbit links 5 frames on 4 nights (0.42--1.38$''$ from its prediction). Strips shifted by
$\pm 60''$ link at most 2 frames at either radius. The preset refit test ($\chi^2$/dof $< 3$) was replaced by the joint orbit fit
that includes the five DECam positions (Table~\ref{tab:chain}, step 4). The group later grew to 7 frames on 5 nights: the $r$ frames of
2018 September 6 and October 17 were measured at the predicted place (a later search on the October 17 frame's own CCD recovers it
0.13$''$ from the orbit). The false-link Monte Carlo of Section~\ref{sec:identity} was run on the 6 frames that were candidates of the
search or of that later search (5 nights); the September 6 $r$ frame is not an independent detection.

\textbf{Stack searches.} Six stack searches of steps 2--5 of Table~\ref{tab:chain} missed one of their preset acceptance rules.
Five are, by those rules, not detections (unconfirmed candidates): after the discovery nights in 2019 one shifted null stack reached S/N 4.15
against a limit of 4; in 2020, 2021 and 2022 one null strip reached 5.52, 5.62 and 5.66 against 5.5; and in 2024 the
static-source test could not be made. In the sixth, the 2019 window of August--September, the shifted null stacks reached S/N 8.1
against a limit of 4.5, so that window was counted as not found. No window or season stack enters the orbit.

\textbf{Orbit class.} The side-by-side test of Section~\ref{sec:discovery} ran the follow-up stage of our pipeline, a different
code from the stack searches of steps 2--5 of Table~\ref{tab:chain}, from 2019 October 26 alone; its free hypothesis has no
resonance and prefers eccentricities near 0 ($\sigma_e = 0.1$) and distances of 30--70~au. In the two 2019 windows only the Trojan
hypothesis was found, before and after the discovery nights (S/N 19.5 and 14.0, against at most 4.2 and 5.6 in 40 null stacks
each). The 3:2 hypothesis gave a weak signal before the nights (6.0 against a null maximum of 5.3) whose position lay 1.11$''$ from
the joint fit (limit 1$''$); the 2:1 hypothesis was not found; the free hypothesis was not found after the nights, and its
one marginal hit before them (6.34 against 6.14) is the Trojan's own light where the two tracks overlap. Only the Trojan hypothesis
was then searched in the seasons (those of the free hypothesis were too large to search, and in 2026 too shallow, and the 3:2 and 2:1 hypotheses,
without an accepted hit in their own windows, were not run): it was found in 2018, 2020 and 2021 (S/N 23.6, 24.6 and 17.1) and, after
re-planning on the Trojan solution, in 2022--2025 (19.2, 17.8, 18.6 and 11.4). The 2026 season was not stacked: no ZTF exposure on
the path was listed when this search was run.

\textbf{Static sky and scan motion.} SDSS's own catalog lists a source at the predicted place in all three scans; the deeper
Legacy Surveys DR10 catalog has no source within 6$''$ of the three SDSS positions (its $r$ depth there is 24.1--24.3). Within the
2009 September 26 scan the source moves as predicted between its $r$, $i$ and $g$ frames (the summed squares of the normalized $r \rightarrow i$ and $r \rightarrow g$ shifts, which share the $r$ position, are 1.9 with
the predicted motion against 23.9 for a fixed source); the 2008 scan leans the same way (0.3 against 2.6), and 2009 January is
undecided. Of the 23
PS1 detection spots, 17 have no PS1 catalog object within 1.5$''$ with a detection on any other date; the other 6 have one or two.
The rows near those 6 are from single nights (2010 August 17, two in $z$; 2013 November 6, one in $g$; 2011 September 11, one in
$i$ at the edge of a masked detector area), and forced photometry on 3--13 PS1 warps of other dates in the detection's filter finds
nothing at any of the 23 spots (combined S/N $-2.2$ to $+1.3$, where a static source as bright as the detection would give 4--26;
single warps $-3.0$ to $+2.7$) nor at three of the four rows' places ($-1.4$ to $+1.1$, in $r$ and $i$). PS1's mask flags a possible optical ghost at the object's place in the four exposures of 2014 November 28; the detections there move at the predicted rate, and forced photometry finds nothing at the place on other dates.

\textbf{Known objects.} The check along the orbit used local copies of the Minor Planet Center's numbered- and unnumbered-object
observation files of 2026 June 3 (which reach 2026 April 14), its Isolated Tracklet File of 2026 November 8 with every line of our copies since 2026 October 12 that has left it, its comet observations of 2026 September 11, and the observation records of 11,231 further objects, so observations that the Minor Planet Center identified with known objects between 2026 April 14 and October 12 were covered only for those 11,231 objects.
Of the 25,890 orbits with $a > 5$~au in its orbit catalog of 2026 November 8, propagated two-body, the nearest to the object's
position at eight dates of 2010--2026 is 0.49\arcdeg{} away. A repeat of the check of the PS1 detections with the copies of the credit check (the same observation files of 2026 June 3, the
Isolated Tracklet File of 2026 November 8 with its departed lines to November 9, and the observation records of known objects) finds
no observation within 300$''$ and 1~day of any of the 23 (the first check, on the observation files of 2026 June 3 and the
Isolated Tracklet File of 2026 November 3, held 22 of them).

\bibliography{references}{}

\begin{thebibliography}{}
\expandafter\ifx\csname natexlab\endcsname\relax\def\natexlab#1{#1}\fi
\providecommand{\url}[1]{\href{#1}{#1}}
\providecommand{\dodoi}[1]{doi:~\href{http://doi.org/#1}{\nolinkurl{#1}}}
\providecommand{\doeprint}[1]{\href{http://ascl.net/#1}{\nolinkurl{http://ascl.net/#1}}}
\providecommand{\doarXiv}[1]{\href{https://arxiv.org/abs/#1}{\nolinkurl{https://arxiv.org/abs/#1}}}

\bibitem[{ {Abdurro'uf} {et~al.}(2022){Abdurro'uf}, Accetta, Aerts,
  {et~al.}}]{abdurrouf2022}
{Abdurro'uf}, Accetta, K., Aerts, C., {et~al.} 2022, \bibinfo{title}{The
  Seventeenth Data Release of the {Sloan} {Digital} {Sky} {Surveys}: Complete
  Release of {MaNGA}, {MaStar}, and {APOGEE-2} Data,} \apjs, 259, 35,
  \dodoi{10.3847/1538-4365/ac4414}

\bibitem[{A. Alvarez-Candal {et~al.}(2016)Alvarez-Candal, Pinilla-Alonso,
  Ortiz, Duffard, Morales, Santos-Sanz, Thirouin, \& Silva}]{alvarezcandal2016}
Alvarez-Candal, A., Pinilla-Alonso, N., Ortiz, J.~L., {et~al.} 2016,
  \bibinfo{title}{Absolute magnitudes and phase coefficients of
  trans-{Neptunian} objects,} \aap, 586, A155,
  \dodoi{10.1051/0004-6361/201527161}

\bibitem[{ {Astropy Collaboration} {et~al.}(2022){Astropy Collaboration},
  Price-Whelan, Lim, Earl, {et~al.}}]{astropy2022}
{Astropy Collaboration}, Price-Whelan, A.~M., Lim, P.~L., Earl, N., {et~al.}
  2022, \bibinfo{title}{The {Astropy} Project: Sustaining and Growing a
  Community-oriented Open-source Project and the Latest Major Release (v5.0) of
  the Core Package,} \apj, 935, 167, \dodoi{10.3847/1538-4357/ac7c74}

\bibitem[{ {Astropy Collaboration} {et~al.}(2018){Astropy Collaboration},
  Price-Whelan, Sip{\H{o}}cz, G{\"u}nther, {et~al.}}]{astropy2018}
{Astropy Collaboration}, Price-Whelan, A.~M., Sip{\H{o}}cz, B.~M., G{\"u}nther,
  H.~M., {et~al.} 2018, \bibinfo{title}{The {Astropy} Project: Building an
  Open-science Project and Status of the v2.0 Core Package,} \aj, 156, 123,
  \dodoi{10.3847/1538-3881/aabc4f}

\bibitem[{ {Astropy Collaboration} {et~al.}(2013){Astropy Collaboration},
  Robitaille, Tollerud, Greenfield, {et~al.}}]{astropy2013}
{Astropy Collaboration}, Robitaille, T.~P., Tollerud, E.~J., Greenfield, P.,
  {et~al.} 2013, \bibinfo{title}{Astropy: A community {Python} package for
  astronomy,} \aap, 558, A33, \dodoi{10.1051/0004-6361/201322068}

\bibitem[{D. Baade {et~al.}(1999)Baade, Meisenheimer, Iwert,
  {et~al.}}]{baade1999}
Baade, D., Meisenheimer, K., Iwert, O., {et~al.} 1999, \bibinfo{title}{The
  {Wide} {Field} {Imager} at the 2.2-m {MPG/ESO} telescope: first views with a
  67-million-facette eye,} The Messenger, 95, 15

\bibitem[{E.~C. Bellm {et~al.}(2019)Bellm, Kulkarni, Graham,
  {et~al.}}]{bellm2019}
Bellm, E.~C., Kulkarni, S.~R., Graham, M.~J., {et~al.} 2019,
  \bibinfo{title}{The {Zwicky} {Transient} {Facility}: System Overview,
  Performance, and First Results,} \pasp, 131, 018002,
  \dodoi{10.1088/1538-3873/aaecbe}

\bibitem[{P.~H. Bernardinelli {et~al.}(2022)Bernardinelli, Bernstein, Sako,
  Yanny, Aguena, Allam, {et~al.}}]{bernardinelli2022}
Bernardinelli, P.~H., Bernstein, G.~M., Sako, M., {et~al.} 2022,
  \bibinfo{title}{A Search of the Full Six Years of the {Dark} {Energy}
  {Survey} for Outer {Solar} {System} Objects,} \apjs, 258, 41,
  \dodoi{10.3847/1538-4365/ac3914}

\bibitem[{O. Boulade {et~al.}(2003)Boulade, Charlot, Abbon, Aune, Borgeaud,
  Carton, {et~al.}}]{boulade2003}
Boulade, O., Charlot, X., Abbon, P., {et~al.} 2003, \bibinfo{title}{{MegaCam}:
  the new {Canada-France-Hawaii} {Telescope} wide-field imaging camera,} in
  Proc. SPIE, Vol. 4841, Instrument Design and Performance for Optical/Infrared
  Ground-based Telescopes, 72, \dodoi{10.1117/12.459890}

\bibitem[{E. Bowell {et~al.}(1989)Bowell, Hapke, Domingue, Lumme, Peltoniemi,
  \& Harris}]{bowell1989}
Bowell, E., Hapke, B., Domingue, D., {et~al.} 1989, \bibinfo{title}{Application
  of photometric models to asteroids,} in Asteroids {II}, ed. R.~P. Binzel,
  T.~Gehrels, \& M.~S. Matthews (Tucson: University of Arizona Press), 524--556

\bibitem[{ {Center for Near Earth Object Studies}(2026){Center for Near Earth
  Object Studies}}]{cneossize}
{Center for Near Earth Object Studies}. 2026, Asteroid Size Estimator,
  NASA/JPL, accessed 2026 November 11.
\newblock \url{https://cneos.jpl.nasa.gov/tools/ast_size_est.html}

\bibitem[{G. Cerati(2026)Cerati}]{cerati2026}
Cerati, G. 2026, An Archival Search for Outer Solar System Objects in the
  {NOIRLab} Source Catalog {DR2}: 65 Candidates, Including Four {L4} {Neptune}
  {Trojans},, arXiv:2611.14330 \doarXiv{2611.14330}

\bibitem[{K.~C. Chambers {et~al.}(2016)Chambers, Magnier, Metcalfe, Flewelling,
  Huber, {et~al.}}]{chambers2016}
Chambers, K.~C., Magnier, E.~A., Metcalfe, N., {et~al.} 2016, The {Pan-STARRS1}
  Surveys, \doarXiv{1612.05560}

\bibitem[{S.~R. Chesley {et~al.}(2017)Chesley, Hockney, \&
  Holman}]{chesley2017}
Chesley, S.~R., Hockney, G.~M., \& Holman, M.~J. 2017,
  \bibinfo{title}{Introducing {ADES}: A New {IAU} Astrometry Data Exchange
  Standard,} in AAS/Division for Planetary Sciences Meeting Abstracts, Vol.~49,
  112.14

\bibitem[{E.~I. Chiang {et~al.}(2003)Chiang, Jordan, Millis, Buie, Wasserman,
  Elliot, Kern, Trilling, Meech, \& Wagner}]{chiang2003}
Chiang, E.~I., Jordan, A.~B., Millis, R.~L., {et~al.} 2003,
  \bibinfo{title}{Resonance Occupation in the {Kuiper} Belt: Case Examples of
  the 5:2 and {Trojan} Resonances,} \aj, 126, 430, \dodoi{10.1086/375207}

\bibitem[{R. Dekany {et~al.}(2020)Dekany, Smith, Riddle, Feeney,
  {et~al.}}]{dekany2020}
Dekany, R., Smith, R.~M., Riddle, R., Feeney, M., {et~al.} 2020,
  \bibinfo{title}{The {Zwicky} {Transient} {Facility}: Observing System,}
  \pasp, 132, 038001, \dodoi{10.1088/1538-3873/ab4ca2}

\bibitem[{R. Dvorak {et~al.}(2007)Dvorak, Schwarz, S{\"u}li, \&
  Kotoulas}]{dvorak2007}
Dvorak, R., Schwarz, R., S{\"u}li, {\'A}., \& Kotoulas, T. 2007,
  \bibinfo{title}{On the stability of the {Neptune} {Trojans},} \mnras, 382,
  1324, \dodoi{10.1111/j.1365-2966.2007.12480.x}

\bibitem[{D.~J. Eisenstein {et~al.}(2011)Eisenstein, Weinberg, Agol, Aihara,
  {et~al.}}]{eisenstein2011}
Eisenstein, D.~J., Weinberg, D.~H., Agol, E., Aihara, H., {et~al.} 2011,
  \bibinfo{title}{{SDSS-III}: Massive Spectroscopic Surveys of the Distant
  Universe, the {Milky} {Way}, and Extra-Solar Planetary Systems,} \aj, 142,
  72, \dodoi{10.1088/0004-6256/142/3/72}

\bibitem[{B. Flaugher {et~al.}(2015)Flaugher, Diehl, Honscheid, Abbott,
  {et~al.}}]{flaugher2015}
Flaugher, B., Diehl, H.~T., Honscheid, K., Abbott, T. M.~C., {et~al.} 2015,
  \bibinfo{title}{The {Dark} {Energy} {Camera},} \aj, 150, 150,
  \dodoi{10.1088/0004-6256/150/5/150}

\bibitem[{H.~A. Flewelling {et~al.}(2020)Flewelling, Magnier, Chambers,
  {et~al.}}]{flewelling2020}
Flewelling, H.~A., Magnier, E.~A., Chambers, K.~C., {et~al.} 2020,
  \bibinfo{title}{The {Pan-STARRS1} Database and Data Products,} \apjs, 251, 7,
  \dodoi{10.3847/1538-4365/abb82d}

\bibitem[{ {Gaia Collaboration} {et~al.}(2016){Gaia Collaboration}, Prusti,
  de~Bruijne, Brown, {et~al.}}]{gaia2016}
{Gaia Collaboration}, Prusti, T., de~Bruijne, J. H.~J., Brown, A. G.~A.,
  {et~al.} 2016, \bibinfo{title}{The {Gaia} mission,} \aap, 595, A1,
  \dodoi{10.1051/0004-6361/201629272}

\bibitem[{ {Gaia Collaboration} {et~al.}(2023){Gaia Collaboration}, Vallenari,
  Brown, Prusti, {et~al.}}]{gaia2023}
{Gaia Collaboration}, Vallenari, A., Brown, A. G.~A., Prusti, T., {et~al.}
  2023, \bibinfo{title}{{Gaia} {Data} {Release} 3. Summary of the content and
  survey properties,} \aap, 674, A1, \dodoi{10.1051/0004-6361/202243940}

\bibitem[{D.~W. Gerdes {et~al.}(2016)Gerdes, Jennings, Bernstein, Sako, Adams,
  Goldstein, {et~al.}}]{gerdes2016}
Gerdes, D.~W., Jennings, R.~J., Bernstein, G.~M., {et~al.} 2016,
  \bibinfo{title}{Observation of Two New {L4} {Neptune} {Trojans} in the {Dark}
  {Energy} {Survey} Supernova Fields,} \aj, 151, 39,
  \dodoi{10.3847/0004-6256/151/2/39}

\bibitem[{J.~D. Giorgini \&  {JPL Solar System Dynamics Group}(2026)Giorgini \&
  {JPL Solar System Dynamics Group}}]{horizons}
Giorgini, J.~D., \& {JPL Solar System Dynamics Group}. 2026, {NASA/JPL}
  {Horizons} On-Line Ephemeris System, data retrieved 2026-Oct-25, 2026-Nov-04
  and 2026-Nov-06.
\newblock \url{https://ssd.jpl.nasa.gov/horizons/}

\bibitem[{R. Gomes \& D. Nesvorn{\'y}(2016)Gomes \& Nesvorn{\'y}}]{gomes2016}
Gomes, R., \& Nesvorn{\'y}, D. 2016, \bibinfo{title}{Neptune trojan formation
  during planetary instability and migration,} \aap, 592, A146,
  \dodoi{10.1051/0004-6361/201527757}

\bibitem[{M.~J. Graham {et~al.}(2019)Graham, Kulkarni, Bellm,
  {et~al.}}]{graham2019}
Graham, M.~J., Kulkarni, S.~R., Bellm, E.~C., {et~al.} 2019,
  \bibinfo{title}{The {Zwicky} {Transient} {Facility}: Science Objectives,}
  \pasp, 131, 078001, \dodoi{10.1088/1538-3873/ab006c}

\bibitem[{T. Grav {et~al.}(2011)Grav, Mainzer, Bauer, Masiero, Spahr, McMillan,
  Walker, Cutri, Wright, Eisenhardt, Blauvelt, DeBaun, Elsbury, Gautier,
  Gomillion, Hand, \& Wilkins}]{grav2011}
Grav, T., Mainzer, A.~K., Bauer, J., {et~al.} 2011,
  \bibinfo{title}{{WISE}/{NEOWISE} Observations of the {Jovian} {Trojans}:
  Preliminary Results,} \apj, 742, 40, \dodoi{10.1088/0004-637X/742/1/40}

\bibitem[{B. Gray(2022)Gray}]{findorb}
Gray, B. 2022, {Find\_Orb}: Orbit determination from observations, Astrophysics
  Source Code Library, record ascl:2202.016.
\newblock \url{https://www.projectpluto.com/find_orb.htm}

\bibitem[{J.~E. Gunn {et~al.}(1998)Gunn, Carr, Rockosi, Sekiguchi,
  {et~al.}}]{gunn1998}
Gunn, J.~E., Carr, M., Rockosi, C., Sekiguchi, M., {et~al.} 1998,
  \bibinfo{title}{The {Sloan} {Digital} {Sky} {Survey} Photometric Camera,}
  \aj, 116, 3040, \dodoi{10.1086/300645}

\bibitem[{J.~E. Gunn {et~al.}(2006)Gunn, Siegmund, Mannery, Owen,
  {et~al.}}]{gunn2006}
Gunn, J.~E., Siegmund, W.~A., Mannery, E.~J., Owen, R.~E., {et~al.} 2006,
  \bibinfo{title}{The 2.5 m Telescope of the {Sloan} {Digital} {Sky} {Survey},}
  \aj, 131, 2332, \dodoi{10.1086/500975}

\bibitem[{C.~R. Harris {et~al.}(2020)Harris, Millman, van~der Walt, Gommers,
  {et~al.}}]{numpy2020}
Harris, C.~R., Millman, K.~J., van~der Walt, S.~J., Gommers, R., {et~al.} 2020,
  \bibinfo{title}{Array programming with {NumPy},} \nat, 585, 357,
  \dodoi{10.1038/s41586-020-2649-2}

\bibitem[{J. Horner \& P.~S. Lykawka(2010)Horner \& Lykawka}]{horner2010}
Horner, J., \& Lykawka, P.~S. 2010, \bibinfo{title}{The {Neptune} {Trojans} --
  a new source for the {Centaurs}?} \mnras, 402, 13,
  \dodoi{10.1111/j.1365-2966.2009.15702.x}

\bibitem[{J.~D. Hunter(2007)Hunter}]{matplotlib2007}
Hunter, J.~D. 2007, \bibinfo{title}{Matplotlib: A {2D} Graphics Environment,}
  Computing in Science \& Engineering, 9, 90, \dodoi{10.1109/MCSE.2007.55}

\bibitem[{ {IRSA}(2022){IRSA}}]{ztfimages}
{IRSA}. 2022, {Zwicky} {Transient} {Facility} Image Service, IPAC,
  \dodoi{10.26131/IRSA539}

\bibitem[{{\v Z}. Ivezi{\'c} {et~al.}(2019)Ivezi{\'c}, Kahn, Tyson,
  {et~al.}}]{ivezic2019}
Ivezi{\'c}, {\v Z}., Kahn, S.~M., Tyson, J.~A., {et~al.} 2019,
  \bibinfo{title}{{LSST}: From Science Drivers to Reference Design and
  Anticipated Data Products,} \apj, 873, 111, \dodoi{10.3847/1538-4357/ab042c}

\bibitem[{D. Jewitt(2018)Jewitt}]{jewitt2018}
Jewitt, D. 2018, \bibinfo{title}{The {Trojan} Color Conundrum,} \aj, 155, 56,
  \dodoi{10.3847/1538-3881/aaa1a4}

\bibitem[{M.~S.~P. Kelley {et~al.}(2021)Kelley, Darg, \& Bauer}]{kelley2021}
Kelley, M. S.~P., Darg, D., \& Bauer, J. 2021, \bibinfo{title}{{CATCH}ing
  Near-{Earth} Objects in Archival Survey Data,} in 7th {IAA} Planetary Defense
  Conference.
\newblock \url{https://catch.astro.umd.edu/about}

\bibitem[{P. Lacerda {et~al.}(2014)Lacerda, Fornasier, Lellouch, Kiss,
  Vilenius, Santos-Sanz, Rengel, M{\"u}ller, Stansberry, Duffard, Delsanti, \&
  Guilbert-Lepoutre}]{lacerda2014}
Lacerda, P., Fornasier, S., Lellouch, E., {et~al.} 2014, \bibinfo{title}{The
  Albedo-Color Diversity of Transneptunian Objects,} \apjl, 793, L2,
  \dodoi{10.1088/2041-8205/793/1/L2}

\bibitem[{S. Larson {et~al.}(1998)Larson, Brownlee, Hergenrother, \&
  Spahr}]{css}
Larson, S., Brownlee, J., Hergenrother, C., \& Spahr, T. 1998,
  \bibinfo{title}{The {Catalina} {Sky} {Survey} for {NEOs},} \baas, 30, 12P.14

\bibitem[{H.~W. Lin {et~al.}(2019)Lin, Gerdes, Hamilton, Adams, Bernstein,
  Sako, {et~al.}}]{lin2019}
Lin, H.~W., Gerdes, D.~W., Hamilton, S.~J., {et~al.} 2019,
  \bibinfo{title}{Evidence for color dichotomy in the primordial {Neptunian}
  {Trojan} population,} \icarus, 321, 426, \dodoi{10.1016/j.icarus.2018.12.006}

\bibitem[{H.~W. Lin {et~al.}(2016)Lin, Chen, Holman, Ip, Payne, Lacerda,
  Fraser, Gerdes, Bieryla, Sie, Chen, Burgett, Denneau, Jedicke, Kaiser,
  Magnier, Tonry, Wainscoat, \& Waters}]{lin2016}
Lin, H.~W., Chen, Y.-T., Holman, M.~J., {et~al.} 2016, \bibinfo{title}{The
  {Pan-STARRS} 1 Discoveries of Five New {Neptune} {Trojans},} \aj, 152, 147,
  \dodoi{10.3847/0004-6256/152/5/147}

\bibitem[{H.~W. Lin {et~al.}(2021)Lin, Chen, Volk, Gladman, Murray-Clay,
  Alexandersen, Bannister, Lawler, Ip, Lykawka, Kavelaars, Gwyn, \&
  Petit}]{lin2021}
Lin, H.~W., Chen, Y.-T., Volk, K., {et~al.} 2021, \bibinfo{title}{{OSSOS}: The
  eccentricity and inclination distributions of the stable neptunian
  {Trojans},} \icarus, 361, 114391, \dodoi{10.1016/j.icarus.2021.114391}

\bibitem[{L. Lindegren {et~al.}(2021)Lindegren, Klioner, Hern{\'a}ndez,
  Bombrun, {et~al.}}]{lindegren2021}
Lindegren, L., Klioner, S.~A., Hern{\'a}ndez, J., Bombrun, A., {et~al.} 2021,
  \bibinfo{title}{{Gaia} {Early} {Data} {Release} 3. The astrometric solution,}
  \aap, 649, A2, \dodoi{10.1051/0004-6361/202039709}

\bibitem[{P.~S. Lykawka {et~al.}(2009)Lykawka, Horner, Jones, \&
  Mukai}]{lykawka2009}
Lykawka, P.~S., Horner, J., Jones, B.~W., \& Mukai, T. 2009,
  \bibinfo{title}{Origin and dynamical evolution of {Neptune} {Trojans} -- {I}.
  Formation and planetary migration,} \mnras, 398, 1715,
  \dodoi{10.1111/j.1365-2966.2009.15243.x}

\bibitem[{E.~A. Magnier {et~al.}(2020{\natexlab{a}})Magnier, Chambers,
  Flewelling, {et~al.}}]{magnier2020c}
Magnier, E.~A., Chambers, K.~C., Flewelling, H.~A., {et~al.}
  2020{\natexlab{a}}, \bibinfo{title}{The {Pan-STARRS} Data-processing System,}
  \apjs, 251, 3, \dodoi{10.3847/1538-4365/abb829}

\bibitem[{E.~A. Magnier \& J.-C. Cuillandre(2004)Magnier \&
  Cuillandre}]{magnier2004}
Magnier, E.~A., \& Cuillandre, J.-C. 2004, \bibinfo{title}{The {Elixir} System:
  Data Characterization and Calibration at the {Canada-France-Hawaii}
  {Telescope},} \pasp, 116, 449, \dodoi{10.1086/420756}

\bibitem[{E.~A. Magnier {et~al.}(2020{\natexlab{b}})Magnier, Schlafly,
  Finkbeiner, {et~al.}}]{magnier2020}
Magnier, E.~A., Schlafly, E.~F., Finkbeiner, D.~P., {et~al.}
  2020{\natexlab{b}}, \bibinfo{title}{{Pan-STARRS} Photometric and Astrometric
  Calibration,} \apjs, 251, 6, \dodoi{10.3847/1538-4365/abb82a}

\bibitem[{E.~A. Magnier {et~al.}(2020{\natexlab{c}})Magnier, Sweeney, Chambers,
  {et~al.}}]{magnier2020b}
Magnier, E.~A., Sweeney, W.~E., Chambers, K.~C., {et~al.} 2020{\natexlab{c}},
  \bibinfo{title}{{Pan-STARRS} Pixel Analysis: Source Detection and
  Characterization,} \apjs, 251, 5, \dodoi{10.3847/1538-4365/abb82c}

\bibitem[{L. Markwardt {et~al.}(2023)Markwardt, Lin, Gerdes, \&
  Adams}]{markwardt2023}
Markwardt, L., Lin, H.~W., Gerdes, D., \& Adams, F.~C. 2023,
  \bibinfo{title}{Photometric Survey of {Neptune}'s {Trojan} Asteroids. {I}.
  The Color Distribution,} \psj, 4, 135, \dodoi{10.3847/PSJ/ace528}

\bibitem[{L. Markwardt {et~al.}(2025)Markwardt, Lin, Holler, Gerdes, Adams,
  Malhotra, \& Napier}]{markwardt2025}
Markwardt, L., Lin, H.~W., Holler, B.~J., {et~al.} 2025, \bibinfo{title}{From
  Colors to Spectra and Back Again: First Near-{IR} Spectroscopic Survey of
  {Neptunian} {Trojans},} \psj, 6, 154, \dodoi{10.3847/PSJ/addecd}

\bibitem[{F.~J. Masci {et~al.}(2019)Masci, Laher, Rusholme, Shupe,
  {et~al.}}]{masci2019}
Masci, F.~J., Laher, R.~R., Rusholme, B., Shupe, D.~L., {et~al.} 2019,
  \bibinfo{title}{The {Zwicky} {Transient} {Facility}: Data Processing,
  Products, and Archive,} \pasp, 131, 018003, \dodoi{10.1088/1538-3873/aae8ac}

\bibitem[{ {Minor Planet Center}(2021){Minor Planet Center}}]{mpec2021s63}
{Minor Planet Center}. 2021, {MPEC} 2021-{S63}: Twenty-Eight New {TNOs}, Minor
  Planet Electronic Circular, issued 2021 September 22.
\newblock \url{https://minorplanetcenter.net/mpec/K21/K21S63.html}

\bibitem[{ {Minor Planet Center}(2026{\natexlab{a}}){Minor Planet
  Center}}]{mpcntlist}
{Minor Planet Center}. 2026{\natexlab{a}}, List Of {Neptune} {Trojans}, Minor
  Planet Center, last updated 2026 May 30, accessed 2026 November 7.
\newblock \url{https://minorplanetcenter.net/iau/lists/NeptuneTrojans.html}

\bibitem[{ {Minor Planet Center}(2026{\natexlab{b}}){Minor Planet
  Center}}]{mpcbandconv}
{Minor Planet Center}. 2026{\natexlab{b}}, Band conversion to {V}
  ({BandConversion.txt}), Minor Planet Center, accessed 2026 November 11.
\newblock \url{https://www.minorplanetcenter.net/iau/info/BandConversion.txt}

\bibitem[{D. Nesvorn{\'y} \& D. Vokrouhlick{\'y}(2009)Nesvorn{\'y} \&
  Vokrouhlick{\'y}}]{nesvorny2009}
Nesvorn{\'y}, D., \& Vokrouhlick{\'y}, D. 2009, \bibinfo{title}{Chaotic Capture
  of {Neptune} {Trojans},} \aj, 137, 5003, \dodoi{10.1088/0004-6256/137/6/5003}

\bibitem[{R. Okuta {et~al.}(2017)Okuta, Unno, Nishino, Hido, \&
  Loomis}]{okuta2017}
Okuta, R., Unno, Y., Nishino, D., Hido, S., \& Loomis, C. 2017,
  \bibinfo{title}{{CuPy}: A {NumPy}-Compatible Library for {NVIDIA} {GPU}
  Calculations,} in Proceedings of Workshop on Machine Learning Systems
  (LearningSys) in The Thirty-first Annual Conference on Neural Information
  Processing Systems (NIPS).
\newblock \url{http://learningsys.org/nips17/assets/papers/paper_16.pdf}

\bibitem[{R.~S. Park {et~al.}(2021)Park, Folkner, Williams, \&
  Boggs}]{park2021}
Park, R.~S., Folkner, W.~M., Williams, J.~G., \& Boggs, D.~H. 2021,
  \bibinfo{title}{The {JPL} Planetary and Lunar Ephemerides {DE440} and
  {DE441},} \aj, 161, 105, \dodoi{10.3847/1538-3881/abd414}

\bibitem[{D.~L. Rabinowitz {et~al.}(2007)Rabinowitz, Schaefer, \&
  Tourtellotte}]{rabinowitz2007}
Rabinowitz, D.~L., Schaefer, B.~E., \& Tourtellotte, S.~W. 2007,
  \bibinfo{title}{The Diverse Solar Phase Curves of Distant Icy Bodies. {I}.
  Photometric Observations of 18 {Trans-Neptunian} Objects, 7 {Centaurs}, and
  {Nereid},} \aj, 133, 26, \dodoi{10.1086/508931}

\bibitem[{H. Rein \& S.-F. Liu(2012)Rein \& Liu}]{rein2012}
Rein, H., \& Liu, S.-F. 2012, \bibinfo{title}{{REBOUND}: an open-source
  multi-purpose {N}-body code for collisional dynamics,} \aap, 537, A128,
  \dodoi{10.1051/0004-6361/201118085}

\bibitem[{H. Rein \& D. Tamayo(2015)Rein \& Tamayo}]{rein2015}
Rein, H., \& Tamayo, D. 2015, \bibinfo{title}{{WHFAST}: a fast and unbiased
  implementation of a symplectic {Wisdom-Holman} integrator for long-term
  gravitational simulations,} \mnras, 452, 376, \dodoi{10.1093/mnras/stv1257}

\bibitem[{R.~S. Seaman {et~al.}(2022)Seaman, Christensen, Gibbs, Shelly, \&
  Larson}]{seaman2022}
Seaman, R.~S., Christensen, E.~J., Gibbs, A., Shelly, F.~C., \& Larson, S.
  2022, {Catalina} {Sky} {Survey} Bundle {V1.0},
  urn:nasa:pds:gbo.ast.catalina.survey::1.0, NASA Planetary Data System,
  \dodoi{10.26033/80FQ-DN90}

\bibitem[{S.~S. Sheppard \& C.~A. Trujillo(2006)Sheppard \&
  Trujillo}]{sheppard2006}
Sheppard, S.~S., \& Trujillo, C.~A. 2006, \bibinfo{title}{A Thick Cloud of
  {Neptune} {Trojans} and Their Colors,} Science, 313, 511,
  \dodoi{10.1126/science.1127173}

\bibitem[{S.~S. Sheppard \& C.~A. Trujillo(2010)Sheppard \&
  Trujillo}]{sheppard2010}
Sheppard, S.~S., \& Trujillo, C.~A. 2010, \bibinfo{title}{Detection of a
  Trailing ({L5}) {Neptune} {Trojan},} Science, 329, 1304,
  \dodoi{10.1126/science.1189666}

\bibitem[{J.~L. Tonry {et~al.}(2012)Tonry, Stubbs, Lykke, Doherty,
  {et~al.}}]{tonry2012}
Tonry, J.~L., Stubbs, C.~W., Lykke, K.~R., Doherty, P., {et~al.} 2012,
  \bibinfo{title}{The {Pan-STARRS1} Photometric System,} \apj, 750, 99,
  \dodoi{10.1088/0004-637X/750/2/99}

\bibitem[{F. Valdes {et~al.}(2014)Valdes, Gruendl, \& {DES
  Project}}]{valdes2014}
Valdes, F., Gruendl, R., \& {DES Project}. 2014, \bibinfo{title}{The {DECam}
  Community Pipeline,} in Astronomical Society of the Pacific Conference
  Series, Vol. 485, Astronomical Data Analysis Software and Systems {XXIII},
  ed. N.~Manset \& P.~Forshay, 379

\bibitem[{P. Virtanen {et~al.}(2020)Virtanen, Gommers, Oliphant, Haberland,
  {et~al.}}]{scipy2020}
Virtanen, P., Gommers, R., Oliphant, T.~E., Haberland, M., {et~al.} 2020,
  \bibinfo{title}{{SciPy} 1.0: fundamental algorithms for scientific computing
  in {Python},} Nature Methods, 17, 261, \dodoi{10.1038/s41592-019-0686-2}

\bibitem[{C.~Z. Waters {et~al.}(2020)Waters, Magnier, Price,
  {et~al.}}]{waters2020}
Waters, C.~Z., Magnier, E.~A., Price, P.~A., {et~al.} 2020,
  \bibinfo{title}{{Pan-STARRS} Pixel Processing: Detrending, Warping,
  Stacking,} \apjs, 251, 4, \dodoi{10.3847/1538-4365/abb82b}

\bibitem[{J. Wisdom \& M. Holman(1991)Wisdom \& Holman}]{wisdom1991}
Wisdom, J., \& Holman, M. 1991, \bibinfo{title}{Symplectic maps for the n-body
  problem,} \aj, 102, 1528, \dodoi{10.1086/115978}

\bibitem[{J. Wisdom {et~al.}(1996)Wisdom, Holman, \& Touma}]{wisdom1996}
Wisdom, J., Holman, M., \& Touma, J. 1996, \bibinfo{title}{Symplectic
  Correctors,} in Fields Institute Communications, Vol.~10, Integration
  Algorithms and Classical Mechanics (Providence, RI: American Mathematical
  Society), 217

\bibitem[{D.~G. York {et~al.}(2000)York, Adelman, Anderson, Anderson,
  {et~al.}}]{york2000}
York, D.~G., Adelman, J., Anderson, Jr., J.~E., Anderson, S.~F., {et~al.} 2000,
  \bibinfo{title}{The {Sloan} {Digital} {Sky} {Survey}: Technical Summary,}
  \aj, 120, 1579, \dodoi{10.1086/301513}

\bibitem[{L.-Y. Zhou {et~al.}(2009)Zhou, Dvorak, \& Sun}]{zhou2009}
Zhou, L.-Y., Dvorak, R., \& Sun, Y.-S. 2009, \bibinfo{title}{The dynamics of
  {Neptune} {Trojan} -- {I}. The inclined orbits,} \mnras, 398, 1217,
  \dodoi{10.1111/j.1365-2966.2009.15203.x}

\end{thebibliography}
\bibliographystyle{aasjournalv7}

\end{document}